\PassOptionsToPackage{table}{xcolor}
\documentclass[sigconf,nonacm]{acmart}
\AtBeginDocument{%
  }

\setcopyright{acmlicensed}
\copyrightyear{2025}
\acmYear{2025}
\setcopyright{acmlicensed}\acmConference[MM '25]{Proceedings of the 33rd ACM International Conference on Multimedia}{October 27--31, 2025}{Dublin, Ireland}
\acmBooktitle{Proceedings of the 33rd ACM International Conference on Multimedia (MM '25), October 27--31, 2025, Dublin, Ireland}
\acmDOI{10.1145/3746027.3755648}
\acmISBN{979-8-4007-2035-2/2025/10}

\usepackage{multirow}
\usepackage{hyperref}
\usepackage{graphicx}
\usepackage{tabularx}
\usepackage{booktabs}
\usepackage{algorithm}
\usepackage{algpseudocode}
\usepackage{balance}

\begin{document}
\pagestyle{plain}   
\title{Cross-Modal Retrieval with Cauchy-Schwarz Divergence}

\author{Jiahao Zhang}
\email{zhangjiahao0310@gmail.com}
\orcid{0009-0005-5720-4535}
\affiliation{%
  \institution{The HongKong University of Science and Technology (Guangzhou)}
  \city{Guangzhou, Guangdong}
  \country{China}
}

\author{Wenzhe Yin}
\email{w.yin@uva.nl }
\affiliation{%
  \institution{University of Amsterdam}
  \city{Amsterdam}
  \country{The Netherlands}
}

\author{Shujian Yu}
\authornote{Corresponding author.}
\email{yusj9011@gmail.com}
\orcid{0000-0002-6385-1705}
\affiliation{%
  \institution{Vrije Universiteit Amsterdam}
  \city{Amsterdam}
  \country{The Netherlands}
}


\begin{abstract}
Effective cross-modal retrieval requires robust alignment of heterogeneous data types. Most existing methods focus on bi-modal retrieval tasks and rely on distributional alignment techniques such as Kullback-Leibler divergence, Maximum Mean Discrepancy, and correlation alignment. However, these methods often suffer from critical limitations, including numerical instability, sensitivity to hyperparameters, and their inability to capture the full structure of the underlying distributions. 
In this paper, we introduce the Cauchy-Schwarz (CS) divergence, a hyperparameter-free measure that improves both training stability and retrieval performance. We further propose a novel Generalized CS (GCS) divergence inspired by H\"older's inequality. This extension enables direct alignment of three or more modalities within a unified mathematical framework through a bidirectional circular comparison scheme, eliminating the need for exhaustive pairwise comparisons. 
Extensive experiments on six benchmark datasets demonstrate the effectiveness of our method in both bi-modal and tri-modal retrieval tasks. The code of our CS/GCS divergence is publicly available at \url{https://github.com/JiahaoZhang666/CSD}.
\end{abstract}



\maketitle

\section{Introduction}
Multi-modal learning has become a cornerstone of artificial intelligence research, with a particular focus on understanding and integrating information across diverse modalities, including vision, language, and audio. Cross-modal retrieval (CMR), a central task in the field, aims to retrieve relevant items in one modality (e.g., vision) given a query in another modality (e.g., audio). It has demonstrated significant impacts across a wide range of domains. For example, multimedia search engines benefit from enhanced content discovery capabilities~\cite{khan2024invideo, quan2025vifi}; personalized recommendation systems leverage multi-modal signals to improve user experiences~\cite{liu2024multimodal, yang2025egolife, zhang2025composed}; autonomous systems integrate data from cameras, LiDAR, radar, and depth sensors to improve environmental perception~\cite{cho2025ev, salehi2024flash}; medical diagnosis systems combine imaging data with clinical reports for better decision-making~\cite{ejiga2025text, zheng2025anatomy}; and large-scale video retrieval systems utilize natural language descriptions for efficient indexing and search~\cite{de2018clinically}.
Despite these advances, a fundamental challenge in CMR remains the alignment of feature distributions across modalities. The semantic gap between different modalities with heterogeneous data types creates significant variations in their data distributions, making direct comparisons challenging. 



Current approaches to modality distribution alignment include Kullback-Leibler (KL) divergence~\cite{zhang2018deep, liu2018index, li2024supervised,  zhi2020cross, xu2020joint, yang2024alignment}, Maximum Mean Discrepancy (MMD) \cite{gretton2006kernel,shi2022information, he2022category, kim2023improving}, and Correlation Alignment (CORAL) \cite{sun2016return, xi2025cross, zhang2025enhancing}. KL divergence measures the difference in distribution between modalities with the logarithm of a density ratio, which suffers from numerical instability when the denominator is close to zero. MMD measures the Euclidean distance between the kernel mean embeddings of data from two modalities. Its performance heavily depends on the careful selection of the kernel size, which limits its generalizability across datasets. CORAL efficiently aligns second-order statistics for simplicity but fails to capture higher-order dependencies. Another key limitation of existing distributional alignment methods for CMR is that they are typically restricted to two modalities. When extending to more than two modalities, a common and de facto strategy is to compute all pairwise divergences, which leads to quadratic computational complexity with respect to the number of modalities. This approach becomes inefficient as the number of modalities increases.

In this paper, we introduce the emerging Cauchy-Schwarz (CS) divergence~\cite{principe2000information, yu2025conditional,yu2024cauchy} to the task of CMR. We show that CS divergence offers several key advantages over existing methods such as KL divergence, MMD, and CORAL, which none of these alternatives can provide simultaneously. Specifically, CS divergence is hyperparameter-free, numerically stable, and linearly scalable with respect to the number of modalities, unlike pairwise approaches that have quadratic complexity. In addition to its theoretical strengths, our method integrates CS divergence with the widely used cross-modal projection matching (CMPM) mechanism~\cite{zhang2018deep,xu2020joint,jiang2023cross}, achieving superior retrieval accuracy compared to state-of-the-art (SOTA) approaches for both bi-modal and tri-modal retrieval.

The key contributions of this work are summarized as follows:
\begin{itemize} 
	\item To the best of our knowledge, this is the first attempt to introduce CS divergence for CMR. We demonstrate that CS divergence possesses several desirable properties not found in commonly used divergence-based distributional alignment methods, making it a strong surrogate for the alignment module in existing CMR approaches.
	
	\item We extend CS divergence to multiple distributions using the H\"older inequality~\cite{qiang2011generalizations,finner1992generalization}, enabling simultaneous alignment of three or more modalities within a unified, linearly scalable framework. The proposed bidirectional circular comparison further removes the need for pairwise modality comparisons required by traditional methods.
	
	\item Comprehensive experiments on six benchmark datasets show that our method outperforms existing approaches in standard, zero-shot, and tri-modal retrieval tasks. 
\end{itemize}

\section{Related Work}
\subsection{Cross-Modal Retrieval}

A substantial body of research has focused on cross-modal retrieval (CMR) within the framework of common embedding space learning, which can be broadly categorized into supervised and unsupervised approaches~\cite{wang2025cross}. Supervised methods~\cite{zhang2018deep,zhan2020supervised} leverage labeled multi-modal data to learn explicit mappings between modalities, enhancing retrieval accuracy through direct semantic guidance. In contrast, unsupervised methods~\cite{peng2019cm} operate without labels, often employing self-supervised objectives~\cite{alwassel2020self} or clustering-based techniques~\cite{zhang2018unsupervised}. While more flexible and adaptable, these methods may struggle to capture fine-grained semantic relationships. Another line of hashing-based approaches has been proposed to learn binary codes as embedding features~\cite{bronstein2010data, jiang2017deep, zhang2018deep}. These methods primarily address retrieval efficiency, as binary embeddings enable fast Hamming distance computation and require significantly less storage. In this work, we focus on the supervised cross-modal retrieval setting, where the goal is to learn real-valued embedding features.



Early CMR methods, such as Canonical Correlation Analysis (CCA) \cite{hotelling1992relations, rasiwasia2010new, shu2021scalable} and its nonlinear extension Kernel CCA (KCCA) \cite{bach2002kernel, jia2019semantically}, project multimodal data into a shared space by maximizing embedding correlations, relying on handcrafted features. Shallow learning models addressed this limitation by automatically learning feature mappings. Notable examples include Joint Representation Learning (JRL)~\cite{zhai2013learning}, which learns a sparse linear mapping function for each modality by enforcing that projected embeddings from different modalities with the same labels are close to each other. Another example is Adversarial Cross-Modal Retrieval (ACMR)~\cite{wang2017adversarial}, which adopts adversarial training~\cite{goodfellow2014generative} to learn modality-invariant representations in a shared embedding space, making it difficult for a modality classifier to distinguish between modalities based on the learned representations.
Despite these advantages, shallow models struggled with complex semantic relationships.

Deep learning transformed CMR by enabling end-to-end representation learning. Deep Canonical Correlation Analysis (DCCA) \cite{andrew2013deep} uses deep neural networks to learn complex nonlinear transformations for each modality (or view) of data. Cross-Modal Projection Matching (CMPM) \cite{zhang2018deep} aligns feature distributions using KL divergence and incorporates matching relationships into an auxiliary classification task, further enhancing the representation compactness within each category. Advanced techniques include TANSS \cite{xu2019ternary}, which introduces structured similarity constraints; CM-GANs \cite{peng2019cm}, which leverages Generative Adversarial Networks (GANs) \cite{goodfellow2014generative} to learn joint representations; and DSCMR \cite{zhen2019deep}, which enforces cross-modal semantic consistency through generative modeling.


{Large multimodal foundation models, such as CLIP~\cite{radford2021learning} and CLAP~\cite{wu2023large}, have been widely adopted for various downstream tasks, including zero-shot cross-modal retrieval. However, CLIP suffers from a modality gap induced by the InfoNCE loss~\cite{liang2022mind, yin2025distributional}, which can lead to unreliable performance on certain tasks (retrieval). In contrast, we focus on cross-modal retrieval by explicitly minimizing the distributional distance between modalities.}



\begin{figure*}[htbp]
	\centering
	\includegraphics[width=2.16\columnwidth, trim=0.6cm 29cm 0.6cm 1cm, clip, page=1]{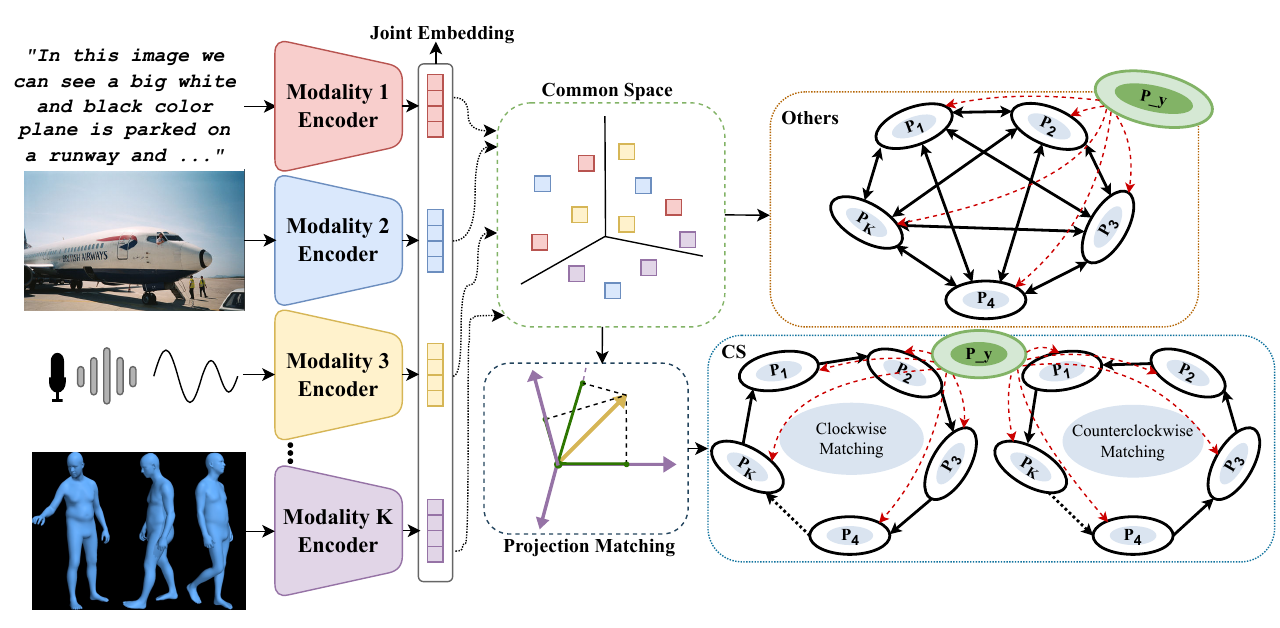} 
	\caption{Illustration of our modality alignment framework with generalized CS divergence. We first construct similarity matrices between each modality and its neighboring modality via projection matching, supervised by the ground-truth matching probabilities derived from the label \( y \). This is followed by a bidirectional circular matching scheme. In contrast, existing methods typically perform exhaustive pairwise comparisons using metrics such as KL divergence or MMD, also guided by the label \( y \).}
	\label{fig:example}
\end{figure*}

\subsection{Feature Distribution Alignment in CMR}

KL divergence is one of the most widely used approaches for feature distribution alignment~\cite{lin2015semantics,zhang2018deep,xu2020joint,chen2021integrating,zhi2020cross,yang2024alignment, li2024supervised}. Typically, given two sets of embeddings from different modalities, $\{\mathbf{e}_i^x\}_{i=1}^{n}$ and $\{\mathbf{e}_j^y\}_{j=1}^{n}$, a similarity matrix $S^{x2y} \in \mathbb{R}^{n \times n}$ is constructed, where the $(i,j)$-th entry represents the similarity between the $i$-th instance in the modality $\mathcal{X}$ and the $j$-th instance in the modality $\mathcal{Y}$. In parallel, a ground-truth similarity matrix $S^{\text{true}}$ is defined, either based on semantic labels or estimated using a pre-trained language model that encodes text labels into embeddings~\cite{yin2024tri}. The KL divergence-based alignment then minimizes the following objective:
\begin{equation}\label{eq:KL_alignment}
\mathcal{L}_{CS} = \sum_{i=1}^n \sum_{j=1}^{n} S^{x2y}_{i, j} \log \frac{S^{x2y}_{i, j}}{S_{i, j}^{\text{true}} + \epsilon}.
\end{equation}


Despite its elegance and effectiveness, KL divergence has a notable limitation: it is sensitive to zero or near-zero probabilities in the denominator, often requiring the addition of an ad hoc smoothing term $\epsilon$. This leads to numerical instability and hyperparameter sensitivity, limiting the robustness and scalability of KL divergence-based alignment in practical retrieval applications.



MMD~\cite{gretton2006kernel} is another widely used approach for feature distribution alignment in CMR~\cite{he2022category, kim2023improving,peng2020unsupervised,jambigi2021mmd}. Formally, MMD computes the Euclidean distance between kernel mean embeddings of $\{\mathbf{e}_i^x\}_{i=1}^{n}$ and $\{\mathbf{e}_j^y\}_{j=1}^{n}$ as follows:
\begin{equation}
	\text{MMD}^2(P(\mathbf{e}^x), P(\mathbf{e}^y)) =
	\| \mathbb{E}_{\mathbf{e}^x}[\phi(\mathbf{e}^x)] - \mathbb{E}_{\mathbf{e}^y}[\phi(\mathbf{e}^y)] \|_{\mathcal{H}}^2,
\end{equation}
where \( \phi(\cdot) \) maps embeddings into a reproducing kernel Hilbert space (RKHS) $\mathcal{H}$. The empirical estimate of MMD is given by:
\begin{equation}
\begin{aligned}
        \widehat{\text{MMD}}^2(P(\mathbf{e}^x), P(\mathbf{e}^y)) & = \frac{1}{n_x^2} \sum_{i,i'=1}^{n_x} \kappa (\mathbf{e}^x_i,\mathbf{e}^x_{i'}) + \frac{1}{n_y^2} \sum_{j,j'=1}^{n_y} \kappa (\mathbf{e}^y_j,\mathbf{e}^y_{j'}) \\
        &- \frac{2}{n_x n_y} \sum_{i,j=1}^{n_x n_y} \kappa (\mathbf{e}^x_i,\mathbf{e}^y_{j}),
\end{aligned}
\end{equation}
where $\kappa$ is a kernel function, typically chosen as the Gaussian kernel $k(\mathbf{u}, \mathbf{v}) = \exp\left(-\frac{\|\mathbf{u} - \mathbf{v}\|_2^2}{2\sigma^2} \right)$ with kernel width $\sigma$. Although MMD captures all higher-order moments of the data distribution~\cite{li2015generative}, its performance is highly sensitive to the choice of $\sigma$. A small $\sigma$ can lead to noise sensitivity, while a large $\sigma$ may blur meaningful structures, thus limiting generalization across diverse datasets.




CORAL~\cite{wang2015unsupervised,xi2025cross,zhang2025enhancing} aligns feature distributions by matching only the second-order statistics of the data:
\begin{equation}
\mathcal{L}_{\text{CORAL}} = \frac{1}{4d^2} \| \mathbf{C}^x - \mathbf{C}^y \|_F^2,
\end{equation}
where \( \| \cdot \|_F \) denotes the Frobenius norm, \( \mathbf{C}_x \) and \( \mathbf{C}_y \) are the sample covariance matrix evaluated over $\{\mathbf{e}_i^x\}_{i=1}^{n}$ and $\{\mathbf{e}_j^y\}_{j=1}^{n}$, respectively.


In addition to the above limitations, it remains unclear how to extend these measures to three or more distributions. As a result, existing approaches in CMR with multiple modalities typically sum all pairwise distances~\cite{zhai2013learning,yin2024tri}, which is computationally inefficient.

\subsection{CS divergence and its Deep Learning Applications}
The Cauchy-Schwarz (CS) divergence, derived from the classic Cauchy-Schwarz inequality, was first introduced in information-theoretic learning \cite{kampa2011closed} and has since been applied in deep clustering \cite{trosten2021reconsidering}, representation learning \cite{yu2024cauchy}, and domain adaptation \cite{yin2024domain}, primarily for pairwise distribution matching in unsupervised settings. 

Despite its stability and scalability, CS divergence remains unexplored in cross-modal retrieval. It has not been extended to multi-modality alignment. We address this gap by introducing CS divergence into supervised retrieval and propose a generalized CS (GCS) divergence based on Hölder's inequality, which enables joint alignment across multiple modalities via circular matching.

\section{Method}

\subsection{Problem Formulation}
Cross-modal retrieval aims to retrieve relevant items from one modality given a query in another modality. Let $\mathcal{O} = \{\mathcal{O}_S, \mathcal{O}_T\}$ represent our cross-modal data consisting of multiple instances (e.g., labeled image-text pairs), where $\mathcal{O}_S$ is the training set and $\mathcal{O}_T$ is the test set. Here, $\mathcal{O}_S = \{(x_i^1, x_i^2, y_i)\}_{i=1}^{N_S}$ and $\mathcal{O}_T = \{(x_j^1, x_j^2, y_j)\}_{j=1}^{N_T}
$ have $N_S$ and $N_T$ instances,
where $x_i^1$ and $x_i^2$ are input samples of the two modalities (e.g., images and text) of the $i$-th instance. Each pair of $\{x_i^1,x_i^2\}$ has been assigned with a semantic label from a pre-defined class set $\mathcal{Y}_S=\{1,2,\cdots,C\}$, where $C$ denotes the total number of classes. In standard retrieval tasks, the class labels are identical for $\mathcal{O}_S$ and $\mathcal{O}_T$, i.e., $\mathcal{Y}_S = \mathcal{Y}_T$. In the zero-shot retrieval, the classes in the two sets are disjoint, I,e, $Y_S \cap Y_T = \emptyset$.

Since different modalities of data typically have different statistical properties and lie in different representation spaces, they cannot be directly compared against each other for cross-modal retrieval~\cite{wang2017adversarial}. Cross-modal learning aims to learn two functions $f$ and $g$ for two modalities: $\mathbf{v}_i=f_\theta(x_i^1)$ and $\mathbf{t}_i=g_\phi(x_i^2)$ that project the input samples from each modality into a common representation space $\mathbb{R}^d$ with dimensionality $d$, where $\theta$ and $\phi$ are trainable parameters of two functions.



This setup can be generalized to more than two modalities. For $M$ modalities, the training and test sets become:
\[
\mathcal{O}_S = \{(x_i^1, x_i^2, ..., x_i^M, y_i)\}_{i=1}^{N_S}, \quad \mathcal{O}_T = \{(x_j^1, x_j^2, ..., x_j^M, y_j)\}_{j=1}^{N_T},
\]
where $x_i^m$ denotes the input sample from the $m$-th modality ($1 \leq m \leq M$) of the $i$-th instance, and $y_i$ be the corresponding semantic label. For each modality, we learn a mapping function that projects the input sample into a common embedding space. A central challenge is how to effectively align the $M$ different modality-specific distributions within a unified framework.



\subsection{Bi-Modal Alignment with CS Divergence}
\label{sec:method-bi-modal}
Without loss of generality, we assume in the following section that the first two modalities are image and text, respectively. Consider a mini-batch with $n$ image-text pairs, where each image feature $\mathbf{v}_i$ is paired with a corresponding text feature $\mathbf{t}_j$. We denote these pairs as $\{(\mathbf{v}_i, \mathbf{t}_j), y_{i,j}\}_{j=1}^n$, where $y_{i,j}$ is a binary indicator: $y_{i,j} = 1$ indicates matched pairs, and $y_{i,j} = 0$ denotes unmatched ones.


The probability $p_{i,j}$ of associating image embedding $\mathbf{v}_i$ with text embedding $\mathbf{t}_j$ can be computed as~\cite{zhang2018deep,xu2020joint,jiang2023cross}:
\begin{equation}
p_{i,j} = \frac{\exp(\cos(\mathbf{v}_i,\mathbf{t}_j))}{\sum_{k=1}^n \exp(\cos(\mathbf{v}_i, \mathbf{t}_k))},
\end{equation}
where $\cos(\cdot,\cdot)$ is the cosine similarity between two embeddings:
\begin{equation}
    \cos(\mathbf{v}_i,\mathbf{t}_j) = \frac{\mathbf{v}_i^\top \mathbf{t}_j}{\|\mathbf{v}_i\|\|\mathbf{t}_j\|}.
\end{equation}


Considering the fact that there might be more than one matched text samples for a single image embedding $\mathbf{v}_i$, the true matching
probability $q_{i,j}$ is normalized as follows:
\begin{equation}
q_{i,j} = \frac{y_{i,j}}{\sum_{k=1}^n y_{i,k}}.
\end{equation}

To align the two discrete probability mass functions (PMFs) $\{p_{i,j}\}_{j=1}^n$ and $\{q_{i,j}\}_{j=1}^n$ for each $\mathbf{v}_i$, where $\sum_{j=1}^n p_{i,j} = 1$ and $\sum_{j=1}^n q_{i,j} = 1$, we are motivated by the famed Cauchy–Schwarz (CS) inequality:
\begin{equation}
\left( \sum_{j=1}^n p_{i,j} q_{i,j} \right)^2 \leq \left( \sum_{j=1}^n p_{i,j}^2 \right) \left( \sum_{j=1}^n q_{i,j}^2 \right),
\end{equation}
with equality if and only if $\{p_{i,j}\}$ and $\{q_{i,j}\}$ are linearly dependent, i.e., proportional or equal to each other. Based on this, the CS divergence between $P_i = [p_{i,1}, p_{i,2}, \dots, p_{i,n}]$ and $Q_i = [q_{i,1}, q_{i,2}, \dots, q_{i,n}]$ is defined as~\cite{jenssen2006cauchy,yu2025conditional}:
\begin{equation}\label{eq:CS_alignment}
	\mathcal{D}_{CS}(P_{i},Q_{i}) = - \log \left( \frac{\sum_{j=1}^{n} p_{i,j} q_{i,j}}{\left( \sum_{j=1}^{n}  p_{i,j}^2 \right)^{\frac{1}{2}} \left( \sum_{j=1}^{n} q_{i,j}^2 \right)^{\frac{1}{2}}} \right).
\end{equation}


A key advantage of CS divergence-based alignment over the KL divergence-based alignment in Eq.~(\ref{eq:KL_alignment}) is its numerical stability. Unlike KL divergence, CS divergence does not suffer from issues related to division by zero and thus avoids the need to introduce a small constant $\epsilon$ to stabilize the denominator. Specifically, by applying the CS inequality again to the vectors $P_i \in \mathbb{R}^n$ (or $Q_i$) and $\mathbf{1} = (1,1,\dots,1) \in \mathbb{R}^n$, we have:
\begin{equation}
\left( \sum_{j=1}^n p_{i,j} \right)^2 \leq \left( \sum_{j=1}^n p_{i,j}^2 \right) \cdot n.
\end{equation}
Since $\sum_{j=1}^n p_{i,j} = 1$, it follows that $\sum_{j=1}^n p_{i,j}^2 \geq \frac{1}{n} $. The same reasoning applies to $\sum_{j=1}^n q_{i,j}^2 \geq \frac{1}{n}$. Therefore, the denominator inside the logarithm in Eq.~(\ref{eq:CS_alignment}) is lower bounded by $1/n$, which ensures numerical stability. This is a big distinction to KL divergence.

Another key advantage of CS divergence-based alignment is that it requires no hyperparameter tuning, unlike methods such as MMD, which depend on selecting an appropriate kernel width $\sigma$.

The loss for image-text matching is computed as:
\begin{equation}
	\mathcal{L}_{v2t}^{\text{true}} = \frac{1}{n}\sum_{i=1}^{n}{D}_{CS}(P_{i},Q_{i}).
\end{equation}

We finally formulate a bidirectional matching loss as:
\begin{equation}\label{eq:loss_CS_alignment}
	\mathcal{L}_{CS} = \mathcal{L}_{v2t}^{\text{true}} + \mathcal{L}_{t2v}^{\text{true}}.
\end{equation}

\subsection{Multi-Modal Alignment with CS Divergence}
\subsubsection{The Generalized Cauchy-Schwarz Divergence} \hfill \\ 
To extend the CS divergence-based alignment to more than two modalities, or even an infinite number, we leverage H\"older's inequality~\cite{qiang2011generalizations,finner1992generalization}  as stated in Theorem~\ref{theorem}.
\begin{theorem}\label{theorem}
Let $K$ be a positive integer. For $M$ sequences of non-negative real numbers $(a_{1,1}, a_{1,2}, \dots, a_{1,K})$, $(a_{2,1}, a_{2,2}, \dots, a_{2,K})$, $\dots$, $(a_{M,1}, a_{M,2}, \dots, a_{M,K})$, 
we have:
\[
\sum_{k=1}^K \prod_{m=1}^M a_{m,k} \leq \prod_{m=1}^M \left( \sum_{k=1}^K a_{m,k}^{M} \right)^{1/M},
\]
with equality if and only if these $M$ sequences are pairwise proportional. The CS inequality is a direct corollary of H\"older inequality when $M=2$.
\end{theorem}

By Theorem~\ref{theorem}, we define the H\"older divergence, also can also be referred to as the generalized Cauchy–Schwarz (GCS) divergence.



\begin{definition}
Let $p_1(x), p_2(x), \dots, p_M(x)$ be $M$ probability mass functions (PMFs) defined over a finite set $\mathcal{X} = \{x_1, x_2, \dots, x_K\}$ with $K$ distinct elements. Let $p_{m,k} = p_m(x_k)$ denote the probability assigned to element $x_k$ by the $m$-th PMF, satisfying $\sum_{k=1}^K p_{m,k} = 1$ for each $m = 1, \dots, M$. The generalized Cauchy–Schwarz divergence, among these $M$ PMFs is defined as:
	\begin{equation}
		\mathcal{D}_{\text{GCS}}(P_1, \dots, P_M) = -\log \left( \frac{\sum_{k=1}^K \prod_{m=1}^M p_{m,k}}{\prod_{m=1}^M \left( \sum_{k=1}^K \left( p_{m,k} \right)^M \right)^{\frac{1}{M}}} \right).
		\label{eq:gcs-discrete}
	\end{equation}
\end{definition}


Eq.~(\ref{eq:gcs-discrete}) defines a valid divergence measure across $M$ discrete distributions. It is symmetric, scale-invariant, and hyperparameter-free\footnote{Proofs of these properties are provided in the supplementary material available in our GitHub repository.}. Similar to the CS divergence, it also avoids the issue of zero denominators, ensuring numerical stability during optimization. This is because the minimum value of $\sum_{k=1}^K p_{m,k}^M$ is $1/K^{M-1}$, which is attained when $p_{m,1} = \dots = p_{m,K} = \frac{1}{K}$.

A key advantage of the GCS divergence is its linear scalability with respect to the number of modalities, making it particularly suitable for multimodal analysis involving more than two, or even a large number of modalities. In cross-modal retrieval, a common approach for handling more than two modalities is to align all pairwise distributions~\cite{yin2024tri,wojcik2024case}, which leads to a computational complexity of $\mathcal{O}(M^2)$. In contrast, GCS scales linearly with $\mathcal{O}(M)$. Compared to the sum of pairwise CS divergences, the proposed GCS formulation captures global inconsistencies that pairwise CS may overlook. While CS compares modality pairs independently, GCS jointly measures divergence across all modalities in a unified and symmetric manner, allowing the detection of subtle mismatches. We refer interested readers to the supplementary material for empirical justification on two synthetic datasets.

\subsubsection{Projection Matching with GCS Divergence}\hfill\\ 
Without loss of generality, we assume three modalities: image, text, and audio. Let $\mathbf{v}_i$, $\mathbf{t}_i$, and $\mathbf{u}_i$ denote the modality-specific embeddings in a shared $d$-dimensional space, learned via their respective modality-specific mapping functions.


Following the CS divergence-based alignment, the probability of associating the $i$-th sample embedding in one modality with the $j$-th sample embedding in another modality in a mini-batch instances of size $n$ can be expressed as:
\begin{equation}
\begin{split}
p_{i,j}^{v2t} & = \frac{\exp(\cos(\mathbf{v}_i,\mathbf{t}_j))}{\sum_{k=1}^n \exp\cos(\mathbf{v}_i, \mathbf{t}_k)}, \\
p_{i,j}^{t2u} & = \frac{\exp(\cos(\mathbf{t}_i,\mathbf{u}_j))}{\sum_{k=1}^n \exp\cos(\mathbf{t}_i, \mathbf{u}_k)}, \\
p_{i,j}^{u2v} & = \frac{\exp(\cos(\mathbf{u}_i,\mathbf{v}_j))}{\sum_{k=1}^n \exp\cos(\mathbf{u}_i, \mathbf{v}_k)},
\end{split}
\end{equation}
where $\cos(\cdot,\cdot)$ refers to the cosine similarity. Similarly, there is a true matching probability $q_{i,j}$ which can be inferred from the given semantic labels.





Therefore, our core idea for multimodal alignment is to align the four PMFs associated with the $i$-th instance, i.e., $P^{v2t}_i=\{p^{v2t}_{i,j}\}_{j=1}^n$, $P^{t2u}_i=\{p^{t2u}_{i,j}\}_{j=1}^n$, $P^{u2v}_i=\{p^{u2v}_{i,j}\}_{j=1}^n$, and $Q_i=\{q_{i,j}\}_{j=1}^n$, using our proposed GCS divergence defined in Eq.~(\ref{eq:gcs-discrete}) as follows:
\begin{equation}
\begin{aligned}
& \mathcal{D}_{GCS}(P_{i}^{v2t}, P_{i}^{t2u}, P_{i}^{u2v}, Q_{i}) 
		 = -\log \left( \sum_{j=1}^{n} p_{i,j}^{v2t} p_{i,j}^{t2u} p_{i,j}^{u2v} q_{i,j} \right) \\
& + \log \left( \sum_{j=1}^{n} \left( p_{i,j}^{v2t} \right)^4 \right)^{\frac{1}{4}} + \log \left( \sum_{j=1}^{n} \left( p_{i,j}^{t2u} \right)^4 \right)^{\frac{1}{4}} \\
& + \log \left( \sum_{j=1}^{n} \left( p_{i,j}^{u2v} \right)^4 \right)^{\frac{1}{4}} + \log \left( \sum_{j=1}^{n} \left( q_{i,j} \right)^4 \right)^{\frac{1}{4}}.          
\end{aligned}
\end{equation}

As in the CS divergence-based alignment, we adopt a bidirectional matching scheme, and the final loss is defined as:
\begin{equation}\label{eq:loss_GCS_alignment}
		\begin{aligned}
		\mathcal{L}_{\text{multi-modal}} &= \frac{1}{n} \sum_{i=1}^{n} \bigg(  \mathcal{D}_{GCS}(P_{i}^{v2t}, P_{i}^{t2u}, P_{i}^{u2v}, Q_{i}) \\
		& + \mathcal{D}_{GCS}(P_{i}^{t2v}, P_{i}^{v2u}, P_{i}^{u2t}, Q_{i}) \bigg).
	\end{aligned}
\end{equation} 

This mechanism can be generalized to an arbitrary number of modalities, as illustrated in the bottom-right panel of Figure~\ref{fig:example}. The forward path aligns the $M$ modalities in a clockwise manner, following the index order $1 \rightarrow 2 \rightarrow \dots \rightarrow M$, while the backward path aligns them counterclockwise, following the order $M \rightarrow (M-1) \rightarrow \dots \rightarrow 1$. Note that, a key distinction from KL divergence and similar methods is that our GCS divergence is \emph{symmetric}, making it \emph{permutation-invariant} with respect to the order of modalities. This symmetry explains why GCS requires only circular comparisons, whereas other methods rely on exhaustive pairwise comparisons.




\subsection{Cross-Modal Retrieval with CS Divergence}
CS divergence and GCS divergence provide an elegant and flexible framework for multimodal alignment that can be seamlessly integrated into state-of-the-art (SOTA) deep cross-modal retrieval methods by replacing KL divergence or MMD-based alignment modules, without requiring any additional modifications, thereby improving both training stability and retrieval performance.

In this paper, we build our bi-modal retrieval method upon Joint Feature Synthesis and Embedding (JFSE)~\cite{xu2020joint}, a SOTA cross-modal retrieval framework that integrates multimodal feature synthesis using class-conditional Wasserstein GANs (cWGANs)~\cite{arjovsky2017wasserstein,felix2018multi} and common embedding space learning within a unified framework. Specifically, each modality is equipped with a cWGAN to generate modality-specific features, which are then transformed into embeddings through corresponding regressors. The original framework performs distributional alignment using either KL divergence or MMD over both real and synthetic embeddings. In our approach, we replace the KL divergence or MMD-based alignment module with the CS divergence objective in Eq.~(\ref{eq:loss_CS_alignment}).

For tri-modal retrieval, we base our method on the LAnguage-VIdeo-MOtion (LAVIMO) alignment framework~\cite{yin2024tri}, which performs distributional alignment in a shared embedding space by minimizing the sum of all pairwise KL divergences. To further enhance alignment, LAVIMO introduces a motion modality reconstruction module, where a decoder takes as input the output of a specially designed attention mechanism. In this mechanism, the motion embedding serves as a query to retrieve relevant information from both the text and video modalities. In our approach, we replace the sum of pairwise KL divergences with the GCS divergence objective in Eq.~(\ref{eq:loss_GCS_alignment}). To demonstrate the plug-and-play flexibility of our method, we apply GCS to other retrieval frameworks. One example is DRCL~\cite{pu2025deep}, which enforces cross-modal consistency via adversarial training and reversible mappings. We replace its original pairwise alignment losses with our GCS divergence, enabling unified distribution-level alignment across modalities while preserving the overall architecture.

We further evaluate the versatility of GCS by integrating it into RONO~\cite{feng2023rono} and COXI~\cite{wei2024cross}. Results for these experiments are provided in the supplementary material.



\section{Experiment}
\subsection{Implementation Details}
\subsubsection{\textbf{Datasets}}\hfill\\
We evaluate our method on \textbf{six} established cross-modal retrieval datasets across three primary tasks. For both standard and zero-shot retrieval tasks, we use \textbf{CUHK-PEDS} \cite{li2017person}\footnote{\url{https://huggingface.co/datasets/PeterPanTheGenius/CUHK-PEDES}} and \textbf{Wikipedia} \cite{rasiwasia2010new}\footnote{\url{https://huggingface.co/datasets/wikimedia/wikipedia}}. \textbf{PKU-XMediaNet} \cite{huang2018mhtn}\footnote{\url{https://www.icst.pku.edu.cn/mipl/lwsjjxz/dataset/index.htm\#XMediaNet}} and \textbf{NUS-WIDE} \cite{chua2009nus}\footnote{\url{https://www.kaggle.com/datasets/xinleili/nuswide}} are specifically used for zero-shot retrieval experiments. For tri-modal retrieval evaluation, we incorporate \textbf{Flickr30k}~\cite{young2014image}~\footnote{\url{https://google.github.io/localized-narratives/index.html}} and \textbf{KIT-ML}~\cite{plappert2016kit}~\footnote{\url{https://motion-annotation.humanoids.kit.edu/dataset/}}.


In all experimental settings, we follow standard evaluation protocols~\cite{rasiwasia2010new, yan2015deep, peng2017ccl}. \textbf{Table}~\ref{tab:data-setting} provides a comprehensive overview of the statistics of the dataset and experimental configurations.
\setlength{\tabcolsep}{2.2pt}
\begin{table}[h]
	\small 
	\caption{The General Statistics of Datasets Used in Standard, Zero-Shot, and Tri-Modal Retrieval Scenarios.}
        \begin{tabularx}{\linewidth}{c|c|c|c|c|c}
		\toprule
		 \multicolumn{1}{c|}{Scenario} & \multicolumn{1}{c|}{Datasets} & \multicolumn{1}{c|}{Modalities}&  \multicolumn{1}{c|}{Train} &  \multicolumn{1}{c|}{Test} &  \multicolumn{1}{c}{Class} \\
		\midrule
		\multirow{2}{*}{Standard} & CUHK-PEDS &\multicolumn{1}{c|}{\multirow{2}{*}{Image-Text}}& 34,405 & 3,074 & 7 \\
		& Wikipedia && 2,173 & 462 & 10 \\
		\midrule
		\multirow{4}{*}{Zero-Shot} & CUHK-PEDS &\multicolumn{1}{c|}{\multirow{4}{*}{Image-Text}}& 34,405 & 4,074 & 4/3 \\
		& Wikipedia && 2,173 & 693 & 5/5 \\
		& NUS-WIDE && 3,894 & 973 & 11/10 \\
		& PKU-XMediaNet && 32000 & 4000 & 100/100 \\
		\midrule
		\multirow{2}{*}{Tri-modal} & Flickr30k &Image-Text-Audio& 30546 & 1023 & 30546 \\
		& KIT-ML &Video-Text-Motion& 3100 & 811 & 55 \\
		\bottomrule
	\end{tabularx}
        \label{tab:data-setting}
\end{table}


\subsubsection{Experimental \textbf{Setup} and Evaluation \textbf{Metrics}}\hfill\\
All experiments are conducted on dual NVIDIA\textsuperscript{\textregistered} RTX 3090 GPUs. We use the Adam optimizer with an initial learning rate of $\mathbf{1 \times 10^{-4}}$, decayed by $\mathbf{0.1}$ every $\mathbf{100}$ epochs, and set $\beta_1 = \mathbf{0.9}$ and $\beta_2 = \mathbf{0.999}$. Models are trained for up to 100 epochs with early stopping based on validation. For optimization stability, we apply gradient clipping (threshold $\mathbf{1.0} $) and a weight decay of $\mathbf{1 \times 10^{-5}}$.
Retrieval performance is evaluated using Precision@K (P@K) with $K = \mathbf{1}$ and $\mathbf{10}$. P@1 assesses the accuracy of the top retrieved item, reflecting the model's ability to identify the most relevant result, while P@10 evaluates performance over a broader candidate set. We report the average performance across $\mathbf{10}$ runs with different random seeds. Hyperparameters are selected via 5-fold cross-validation.

\subsection{Two-Modality Cross-Modal Retrieval}

\subsubsection{\textbf{Standard} Retrieval Evaluation} \hfill \\ 
We evaluate our method against eight state-of-the-art cross-modal retrieval approaches: \textbf{CCA} \cite{rasiwasia2010new}, \textbf{CFA} \cite{li2003multimedia}, \textbf{KCCA} \cite{ballan2014cross}, \textbf{DCCA} \cite{yan2015deep}, \textbf{CM-GANS} \cite{peng2019cm}, \textbf{DSCMR} \cite{zhen2019deep}, \textbf{CMPM} \cite{zhang2018deep} and \textbf{JFSE} \cite{xu2020joint}. Among these, JFSE is particularly relevant to ours, as it uses KL divergence for distributional alignment, while ours replaces KL divergence with CS divergence. To ensure a fair comparison, we use identical image and text feature extractors for both JFSE and our method.  



\setlength{\tabcolsep}{4.5pt}
\begin{table}[h]
	\centering
	\caption{The MAP scores of standard retrieval for Ours and other baselines on CUHK-PEDS and Wikipedia datasets (P@K, K=1). Bolded numbers denote the best performance, and underlined numbers indicate the second-best.}
	\label{tab:standard}
	\begin{tabular}{c|ccc|ccc}
		\toprule
		\multirow{2}{*}{Methods} & \multicolumn{3}{c|}{CUHK-PEDS} & \multicolumn{3}{c}{Wikipedia} \\
		\cmidrule(l{7.4em}r{0.4em}){1-4} \cmidrule(l{3.7em}r{0em}){4-7}
		& I2T & T2I & Avg. & I2T & T2I & Avg. \\
		\midrule
		CCA\cite{rasiwasia2010new}&0.207&0.196&0.202&0.298&0.273&0.286\\
		CFA\cite{li2003multimedia}&0.223&0.207&0.215&0.319&0.316&0.318\\
		KCCA\cite{ballan2014cross}&0.309&0.322&0.316&0.398&0.389&0.394\\
		DCCA\cite{yan2015deep}&0.322&0.347&0.335&0.385&0.379&0.382\\
		DSCMR\cite{zhen2019deep}&0.351&0.372&0.362&0.505&0.471&0.488\\	
		CM-GANS\cite{peng2019cm}&0.427&0.395&0.411&0.507&0.478&0.493\\	
		CMPM\cite{zhang2018deep}&0.411&0.409&0.410&0.494&\underline{0.501}&0.498\\		
		\midrule
		JFSE\cite{xu2020joint}&\underline{0.441}&\underline{0.424}&\underline{0.433}&\underline{0.517}&0.485&\underline{0.501}\\
        \rowcolor{gray!25}
		\bfseries Ours&\bfseries0.467&\bfseries0.434&\bfseries0.451&\bfseries0.533&\bfseries0.507&\bfseries0.520\\
		\bottomrule
	\end{tabular}
\end{table}

\textbf{Table} ~\ref{tab:standard} presents mean average precision
(MAP) scores for standard retrieval on \textbf{CUHK-PEDS} and \textbf{Wikipedia}. Our method consistently outperforms all baseline approaches across both datasets and retrieval directions. On CUHK-PEDS, it achieves the highest average score ($\mathbf{0.449}$), surpassing JFSE ($\mathbf{0.433}$) and other methods like CMPM ($\mathbf{0.410}$) and CM-GANS ($\mathbf{0.411}$). On Wikipedia, our method also leads with $\mathbf{0.520}$, outperforming JFSE ($\mathbf{0.501}$) and other strong baselines. These results validate the effectiveness of CS divergence-based alignment, demonstrating clear gains over both classical (e.g., CCA) and deep learning methods (e.g., DSCMR, CM-GANS).

\subsubsection{\textbf{Zero-Shot} Retrieval Evaluation} \hfill \\
We evaluate our CS divergence-based approach for zero-shot cross-modal retrieval on four benchmarks: \textbf{CUHK-PEDS}, \textbf{Wikipedia}, \textbf{NUS-WIDE}, and \textbf{PKU-XMediaNet}. Following standard protocol, each dataset is split into seen and unseen classes, training exclusively on the seen classes and evaluating on both to assess generalization to novel concepts.
Our method is compared against six established baselines, \textbf{CCA}~\cite{rasiwasia2010new}, \textbf{JRL}~\cite{xu2020joint}, \textbf{KCCA}~\cite{ballan2014cross}, \textbf{DCCA}~\cite{yan2015deep}, \textbf{ACMR}~\cite{zhen2019deep}, and \textbf{TANSS}~\cite{zhang2018deep}, as well as our direct baseline, \textbf{JFSE}~\cite{xu2020joint}. 

\begin{table*}[h]\setlength{\tabcolsep}{8.3pt}
	\begin{center}
		\caption{Quantitative comparison of zero-shot retrieval on \textit{seen classes} of four datasets. The MAP score (P@K, K=1) is reported. Bolded numbers denote the best performance, and underlined numbers indicate the second-best.}
		\label{tab:zs-seen}
		\begin{tabular}{c|lcc|ccc|ccc|ccc}
			\toprule
			\multirow{2}{*}{Methods} & \multicolumn{3}{c|}{CUHK-PEDS} & \multicolumn{3}{c|}{Wikipedia} & \multicolumn{3}{c|}{NUS-WIDE} & \multicolumn{3}{c}{PKU-XMediaNet} \\
			\cmidrule(lr){2-4} \cmidrule(lr){5-7} \cmidrule(lr){8-10} \cmidrule(lr{0em}){11-13} 
			& I2T & T2I & Avg. & I2T & T2I & Avg.& I2T & T2I & Avg. & I2T & T2I & Avg. \\
			\midrule
			CCA\cite{rasiwasia2010new}&0.203&0.211&0.207&0.254&0.265&0.260&0.420&0.435&0.428&0.198&0.253&0.226\\
			JRL\cite{zhai2013learning}&0.435&0.473&0.454&0.517&0.596&0.557&0.473&0.606&0.540&0.329&0.234&0.282\\
			KCCA\cite{ballan2014cross}&0.316&0.335&0.326&0.419&0.517&0.468&0.423&0.476&0.450&0.285&0.323&0.304\\
			DCCA\cite{yan2015deep}&0.352&0.374&0.363&0.447&0.438&0.443&0.425&0.431&0.428&0.213&0.219&0.216\\
			ACMR\cite{wang2017adversarial}&0.475&0.537&0.506&0.651&0.849&0.750&0.597&0.684&0.641&0.697&0.648&0.673\\
			TANSS\cite{xu2019ternary}&0.538&0.605&0.572&0.681&0.875&0.778&0.726&\underline{0.764}&0.745&0.711&0.724&0.718\\
			\midrule
			JFSE\cite{xu2020joint}&\underline{0.593}&\underline{0.611}&\underline{0.602}&\underline{0.694}&\bfseries0.882&\underline{0.788}&\underline{0.754}&0.749&\underline{0.752}&\underline{0.729}&\underline{0.737}&\underline{0.733}\\
			\rowcolor{gray!25}
            \bfseries Ours&\bfseries0.602&\bfseries0.617&\bfseries0.610&\bfseries0.705&\underline{0.878}&\bfseries0.792&\bfseries0.792&\bfseries0.784&\bfseries0.788&\bfseries0.758&\bfseries0.774&\bfseries0.766\\
			\bottomrule
		\end{tabular}
	\end{center}
\end{table*}

\begin{table*}[h]\setlength{\tabcolsep}{8.3pt}
		\centering
		\caption{Quantitative comparison of zero-shot retrieval on \textit{unseen classes} of four datasets. The MAP score (P@K, K=1) is reported. Bolded numbers denote the best performance, and underlined numbers indicate the second-best.} 
		\label{tab:zs-unseen}
		\begin{tabular}{c|lcc|ccc|ccc|ccc}
			\toprule
			\multirow{2}{*}{Methods} & \multicolumn{3}{c|}{CUHK-PEDS} & \multicolumn{3}{c|}{Wikipedia} & \multicolumn{3}{c|}{NUS-WIDE} & \multicolumn{3}{c}{PKU-XMediaNet} \\
			\cmidrule(lr){2-4} \cmidrule(lr){5-7} \cmidrule(lr){8-10} \cmidrule(lr{0em}){11-13} 
			& I2T & T2I & Avg. & I2T & T2I & Avg.& I2T & T2I & Avg. & I2T & T2I & Avg. \\
			\midrule
			CCA\cite{rasiwasia2010new}&0.192&0.161&0.177&0.134&0.118&0.126&0.293&0.274&0.284&0.032&0.041&0.037\\
			JRL\cite{zhai2013learning}&0.209&0.173&0.191&0.175&0.169&0.172&0.346&0.313&0.330&0.038&0.036&0.137\\
			KCCA\cite{ballan2014cross}&0.215&0.174&0.195&0.187&0.173&0.180&0.365&0.319&0.342&0.045&0.051&0.148\\
			DCCA\cite{yan2015deep}&0.237&0.196&0.203&0.210&0.186&0.198&0.372&0.351&0.362&0.063&0.057&0.060\\
			ACMR\cite{wang2017adversarial}&0.275&0.237&0.256&0.251&0.199&0.225&0.417&0.392&0.405&0.035&0.029&0.032\\
			TANSS\cite{xu2019ternary}&0.268&0.231&0.250&0.246&0.227&0.236&0.408&0.384&0.396&\underline{0.128}&0.117&0.123\\
			\midrule
			JFSE\cite{xu2020joint}&\underline{0.384}&\underline{0.437}&\underline{0.411}&\underline{0.340}&\underline{0.312}&\underline{0.326}&\underline{0.492}&\underline{0.503}&\underline{0.498}&0.123&\underline{0.151}&\underline{0.137}\\
			\rowcolor{gray!25}\bfseries Ours&\bfseries0.434&\bfseries0.441&\bfseries0.438&\bfseries0.387&\bfseries0.343&\bfseries0.365&
			\bfseries0.504&\bfseries0.528&\bfseries0.516&\bfseries0.146&\bfseries0.172&\bfseries0.159\\
			\bottomrule
		\end{tabular}
\end{table*}

\textbf{Table}~\ref{tab:zs-seen} shows zero-shot retrieval performance on \textbf{seen} classes across four benchmarks. Our method consistently outperforms all baselines, achieving the highest average \textbf{P@1} scores: $\mathbf{0.610}$ on CUHK-PEDS, $\mathbf{0.792}$ on Wikipedia, $\mathbf{0.788}$ on NUS-WIDE, and $\mathbf{0.766}$ on PKU-XMediaNet.
Compared to JFSE, our method improves by $\mathbf{1.3\%}$ on CUHK-PEDS, $\mathbf{0.5\%}$ on Wikipedia, $\mathbf{4.8\%}$ on NUS-WIDE, and $\mathbf{4.5\%}$ on PKU-XMediaNet. These consistent performance gains across diverse datasets highlight the superiority of CS divergence over KL divergence in cross-modal alignment.

To evaluate generalization, we report zero-shot retrieval on \textbf{unseen} classes. As shown in \textbf{Table}~\ref{tab:zs-unseen}, our method achieves the highest P@1 on all datasets: $\mathbf{0.438}$ (CUHK-PEDS), $\mathbf{0.365}$ (Wikipedia), $\mathbf{0.516}$ (NUS-WIDE), and $\mathbf{0.159}$ (PKU-XMediaNet).
Classical methods like CCA and JRL underperform (e.g., \textbf{0.177} and \textbf{0.191} on CUHK-PEDS), and advanced methods like ACMR and TANSS generalize poorly (0.256 and 0.250).
Compared to JFSE, our method yields gains of $\mathbf{6.6\%}$, $\mathbf{12.0\%}$, $\mathbf{3.6\%}$, and $\mathbf{16.1\%}$ on the four datasets, respectively. These gains show that CS divergence enables stronger alignment under distribution shifts, enhancing transfer to unseen categories.

Interested readers can refer to the supplementary material to visualize latent embeddings (after t-SNE~\cite{van2008visualizing} projection), including image-text feature distributions before and after alignment using KL, MMD, CORAL, and our proposed CS divergence.

\subsection{Three-Modality Cross-Modal Retrieval}
Based on our extension of the CS divergence to three or more modalities via generalized CS divergence, we perform cross-modal retrieval tasks within three modalities.
Specifically, we evaluate our method on two distinct datasets: Flickr30K and KIT-ML, which comprise image-text-audio and video-text-motion triplets, respectively.
We employ the same feature extractor for both image and text modalities. For audio and motion, we utilize wav2vec 2.0~\cite{baevski2020wav2vec} and MotionClip~\cite{clavet2016motion}, respectively. For video inputs, we first extract frame-level representations using CLIP~\cite{radford2021learning}, followed by a temporal modeling step to capture sequential dependencies and enhance video-text alignment.

Note that our baseline, LAnguage-VIdeo-MOtion (\textbf{LAVIMO})~\cite{yin2024tri}, is originally designed for video, text, and motion modalities. To adapt it for the Flickr30K dataset, which involves image, text, and audio modalities, we retain its innovative attention mechanism and pairwise KL divergence-based feature alignment module, but use modality-specific encoders for both images and audio. We refer to this adapted version as LAnguage-IMage-AUdio (\textbf{LAIMAU}). Building upon LAVIMO and LAIMAU, we further propose two variants of our alignment strategy that incorporate CS divergence for tri-modal retrieval. The first variant simply replaces KL divergence with CS divergence while still performing pairwise distributional comparisons for alignment. The second variant eliminates the need for exhaustive pairwise comparisons by directly applying our proposed bidirectional circular comparison, coupled with the generalized CS divergence.
All methods share identical encoder architectures to isolate the effect of the alignment strategies.

\begin{table}[htbp]\setlength{\tabcolsep}{1.2pt}
	\small 
	\caption{Tri-modal retrieval performance on Flikcer30K Datasets. The MAP scores (P@K, K=10) of our methods (CS and GCS) and baselines are reported.}
	\label{tab:LAIMAUA}
	\begin{tabular}{c|ccc|ccc|ccc}
		\toprule
		\multirow{2}{*}{Method} & \multicolumn{9}{c}{Flikcer30K} \\
		\cmidrule(l{6.5em}r{0.3em}){1-4}  \cmidrule(l{2.9em}r{2.9em}){4-8}  \cmidrule(l{2.9em}r{0em}){7-10}  
		& I2T & T2I & Avg.& I2A & A2I & Avg. & A2T & T2A & Avg.  \\
		\midrule
		CCA\cite{rasiwasia2010new}&0.283&0.265&0.274&0.098&0.076&0.087&0.457&0.495&0.476\\
		CFA\cite{li2003multimedia}&0.305&0.293&0.299&0.114&0.128&0.121&0.501&0.482&0.492\\
		KCCA\cite{ballan2014cross}&0.334&0.357&0.346&0.125&0.106&0.116&0.498&0.511&0.505\\
		DCCA\cite{yan2015deep}&0.396&0.378&0.387&0.139&0.131&0.135&0.564&0.571&0.568\\
		CM-GANS\cite{peng2019cm}&0.436&0.477&0.457&0.142&0.127&0.135&0.605&0.621&0.613\\
		DSCMR\cite{zhen2019deep}&0.431&0.428&0.430&0.134&0.116&0.125&0.583&0.597&0.590\\
		CMPM\cite{zhang2018deep}&0.590&0.654&0.622&0.227&0.212&0.220&0.914&0.914&0.914\\
        
        Clip\cite{radford2021learning} &0.605&0.672&0.639&0.148&0.131&0.140&0.809&0.856&0.833\\
        \midrule
        DRCL\cite{pu2025deep}&0.695&0.707&0.701&0.317&0.331&0.322&0.959&0.856&0.908\\
        \rowcolor{gray!25} DRCL+\bfseries CS&\underline{0.759}&\underline{0.726}&\underline{0.743}&\underline{0.332}&0.339&0.336&0.970&0.916&0.943\\ 
        \rowcolor{gray!25} DRCL+\bfseries GCS&0.728&\bfseries0.729&0.729&\bfseries0.339&\underline{0.345}&\bfseries0.342&\underline{0.982}&0.891&0.937\\
        \midrule
		LAIMAU\cite{yin2024tri}&0.692&0.667&0.680&0.288&0.267&0.278&0.964&0.990&0.977\\
        \rowcolor{gray!25}	
        LAIMAU+\bfseries CS&\bfseries0.795&0.705&0.750&0.305&0.328&0.317&\bfseries0.985&\underline{0.993}&\bfseries0.989\\
	  \rowcolor{gray!25}
        LAIMAU+\bfseries GCS&0.705&0.712&0.709&0.323&\bfseries0.358&\underline{0.341}&0.971&\bfseries0.997&\underline{0.984}\\
		\bottomrule
	\end{tabular}
\end{table}

\subsubsection{Analysis of Tri-Modal Retrieval Performance on \textbf{Flickr30K}}\hfill \\
\textbf{Table}~\ref{tab:LAIMAUA} reports P@10 scores for tri-modal retrieval on Flickr30K across three modality pairs: image-text (I2T/T2I), image-audio (I2A/A2I), and audio-text (A2T/T2A). Traditional methods (e.g., CCA, DCCA) perform poorly, particularly on image-audio retrieval (e.g., DCCA: \textbf{0.135} avg). Deep learning baselines, such as CMPM and CM-GANS, show moderate improvements, while LAIMAU achieves better overall performance by utilizing KL-based alignment and attention.
Our CS-based variants consistently outperform baselines. LAIMAU+CS yields strong gains in all directions, with P@10 scores of \textbf{0.750} (I2T/T2I), \textbf{0.317} (I2A/A2I), and \textbf{0.989} (A2T/T2A). Notably, LAIMAU+GCS further improves image-audio retrieval to \textbf{0.341}, a +\textbf{22.7\%} increase over LAIMAU. Similarly, DRCL+CS and DRCL+GCS outperform their original counterparts. These results highlight the effectiveness and generalizability of CS and GCS divergence for robust multi-modal alignment.

\begin{table}[htbp]\setlength{\tabcolsep}{1.2pt}

	\small 
	\caption{Tri-modal retrieval performance on KIT-ML Dataset. The MAP scores (P@K, K=10) of our methods (CS and GCS) and baselines are reported. }
	\label{tab:LAVIMO}
	\begin{tabular}{c|ccc|ccc|ccc}
		\toprule
		\multirow{2}{*}{Method} & \multicolumn{9}{c}{KIT-ML} \\
		\cmidrule(l{6.7em}r{0.3em}){1-4}  \cmidrule(l{2.9em}r{3em}){4-8}  \cmidrule(l{2.9em}r{0em}){7-10}  
		& V2T & T2V & Avg.& V2M & M2V & Avg. & M2T & T2M & Avg.  \\
		\midrule
		CCA\cite{rasiwasia2010new}&0.287&0.272&0.280&0.304&0.292&0.298&0.332&0.295&0.314\\
		CFA\cite{li2003multimedia}&0.292&0.263&0.278&0.317&0.283&0.300&0.305&0.287&0.296\\
		KCCA\cite{ballan2014cross}&0.307&0.350&0.329&0.412&0.396&0.404&0.445&0.401&0.425\\
		DCCA\cite{yan2015deep}&0.423&0.403&0.413&0.457&0.420&0.439&0.429&0.473&0.433\\
		CM-GANS\cite{peng2019cm}&0.585&0.531&0.558&0.643&0.627&0.635&0.612&0.627&0.620\\
		DSCMR\cite{zhen2019deep}&0.607&0.691&0.649&0.659&0.635&0.647&0.687&0.679&0.683\\
		CMPM\cite{zhang2018deep}&0.592&0.624&0.608&0.622&0.658&0.640&0.648&0.653&0.651\\
        
        MotionClip\cite{tevet2022motionclip} &0.609&0.658&0.634&0.864&0.828&0.846&0.645&0.692&0.669\\
        TMR\cite{petrovich2023tmr}&0.573&0.619&0.596&0.707&0.735&0.721&0.919&0.926&0.923\\
		\midrule
        DRCL\cite{pu2025deep}&0.633&0.551&0.592&0.701&0.730&0.716&0.809&0.766&0.788\\
        \rowcolor{gray!25} DRCL+\bfseries CS&0.678&0.605&0.640&0.747&0.737&0.742&0.842&0.808&0.825\\ 
        \rowcolor{gray!25} DRCL+\bfseries GCS&0.669&0.637&0.653&0.769&0.740&0.755&0.852&0.809&0.831\\
        \midrule
		LAVIMO\cite{yin2024tri}&0.725&0.776&0.751&0.987&0.997&0.992&0.943&0.917&0.930\\
	\rowcolor{gray!25}	
        LAVIMO+\bfseries CS&0.745&0.791&0.768&0.979&\bfseries0.999&\underline{0.989}&\bfseries0.985&\bfseries0.993&\bfseries0.989\\
	\rowcolor{gray!25}	
        LAVIMO+\bfseries GCS&\bfseries0.804&\bfseries0.820&\bfseries0.812&\bfseries{0.991}&\bfseries0.998&\bfseries0.995&\underline{0.954}&\underline{0.932}&\underline{0.943}\\
		\bottomrule
	\end{tabular}
\end{table}

\subsubsection{Analysis of Tri-Modal Retrieval Performance on \textbf{KIT-ML}}\hfill \\
\textbf{Table}~\ref{tab:LAVIMO} presents P@10 scores on the KIT-ML dataset across three retrieval tasks: video-text (V2T/T2V), video-motion (V2M/M2V), and motion-text (M2T/T2M). Traditional methods (e.g., KCCA, DCCA) perform poorly, with average scores below \textbf{0.43}. Deep models like DSCMR improve results (e.g., \textbf{0.649} for V2T/T2V), while LAVIMO further enhances performance via KL-based alignment and attention (\textbf{0.751}, \textbf{0.992}, and \textbf{0.930}).
Our CS-based variants outperform all baselines. LAVIMO+CS improves video-text retrieval to \textbf{0.768}, while maintaining competitive performance in other directions. LAVIMO+GCS achieves the best results: \textbf{0.804} for V2T/T2V (\textbf{+7.1\%}), \textbf{0.998} for V2M/M2V (\textbf{+0.5\%}), and 0.943 for M2T/T2M (\textbf{+1.7\%}). Notably, DRCL also benefits from our divergence, with DRCL+GCS outperforming DRCL+CS across all retrieval paths. These gains confirm the effectiveness of GCS in capturing complex tri-modal relationships via joint distribution alignment.

Interested readers can refer to the supplementary material to visualize latent embeddings (after \textbf{t-SNE}~\cite{van2008visualizing} projection), including image-text distributions aligned using KL, MMD, CORAL, and CS divergence. We also provide tri-modal visualizations for image-text-audio and video-text-motion settings, illustrating the effectiveness of our proposed GCS divergence in aligning three modalities jointly.

\begin{table}[htbp]\setlength{\tabcolsep}{1.2pt}\small
   \newcommand{\tabincell}[2]{\begin{tabular}{@{}#1@{}}#2\end{tabular}}
    \begin{center}
    \caption{The ablation study of the proposed clockwise matching and counterclockwise matching strategies on the Flikcer30K dataset. The MAP score (P@K, K=10) is reported. }
	\label{tab:ablation}
	\begin{tabular}{c|ccc|ccc|ccc}
		\toprule
		\multirow{2}{*}{Method} & \multicolumn{9}{c}{Flikcer30K} \\
		\cmidrule(l{6.7em}r{0.3em}){1-4}  \cmidrule(l{2.9em}r{3em}){4-8}  \cmidrule(l{2.9em}r{0em}){7-10}  
		& I2T & T2I & Avg.& I2A & A2I & Avg. & A2T & T2A & Avg.  \\
		\midrule	 
	\rowcolor{gray!15}\bfseries 
       \tabincell{c}{ Clockwise \\
        (I→T→A)}&0.643&/&0.643&/&0.191&0.191&/&0.805&0.805\\
        \rowcolor{gray!15}\bfseries
        \tabincell{c}{ Counter-\\clockwise\\
        (A→T→I)}&/&0.622&0.622&0.200&/&0.200&\bfseries0.796&/&0.796\\
        \midrule	
		\rowcolor{gray!25} \bfseries \tabincell{c}{Mixed\\Bidirectional}&\bfseries0.705&\bfseries0.712&\bfseries0.709&\bfseries0.323&\bfseries0.358&\bfseries0.341&\bfseries0.971&\bfseries0.997&\bfseries0.984\\
		\bottomrule
	\end{tabular}
    \end{center}
\end{table}

\subsubsection{Ablation Study on Matching Strategy for GCS Divergence}\hfill \\
\textbf{Table}~\ref{tab:ablation} reports the performance of different alignment strategies for the Ours(GCS) model on the Flickr30K dataset across three modalities. The \textbf{clockwise} and \textbf{counterclockwise} variants implement unidirectional matching sequences (e.g., \textbf{I→T→A} or \textbf{A→T→I}), while the \textbf{mixed} version incorporates both directions during training. The mixed strategy achieves the best overall performance, particularly in A2T and T2A tasks, confirming the benefit of bidirectional matching for semantically complex modalities.

We compare three matching strategies under our GCS-based framework. The mixed strategy achieves the highest performance across all tasks, with an overall MAP of \textbf{0.984}, outperforming both unidirectional baselines.
While clockwise matching yields reasonable results in some tasks (e.g., A2T), it underperforms in I2A and A2I. The counterclockwise variant is similarly limited. In contrast, the mixed strategy captures semantic relations in all directions, demonstrating the advantage of bidirectional alignment for robust tri-modal retrieval.

\section{Conclusions and Future Work}
This work introduced the Cauchy-Schwarz (CS) divergence as a stable, hyperparameter-free alternative to the KL and MMD divergences for aligning modality distributions in cross-modal retrieval. Extending CS to multiple modalities via a bidirectional circular strategy enables efficient joint alignment without the need for exhaustive pairwise computations. Extensive experiments across five modalities, including image, text, video, audio, and motion, demonstrate consistent gains over state-of-the-art methods, especially in zero-shot settings. While our method builds upon JFSE and LAVIMO, the CS framework is broadly applicable. It can be integrated into various existing models. As multimodal data becomes increasingly diverse, our generalized CS divergence offers a scalable and effective solution for complex alignment challenges in retrieval and beyond, including domain adaptation and contrastive learning.

\bibliographystyle{format}
\balance
\bibliography{main}


\begin{thebibliography}{87}


\ifx \showCODEN    \undefined \def \showCODEN     #1{\unskip}     \fi
\ifx \showISBNx    \undefined \def \showISBNx     #1{\unskip}     \fi
\ifx \showISBNxiii \undefined \def \showISBNxiii  #1{\unskip}     \fi
\ifx \showISSN     \undefined \def \showISSN      #1{\unskip}     \fi
\ifx \showLCCN     \undefined \def \showLCCN      #1{\unskip}     \fi
\ifx \shownote     \undefined \def \shownote      #1{#1}          \fi
\ifx \showarticletitle \undefined \def \showarticletitle #1{#1}   \fi
\ifx \showURL      \undefined \def \showURL       {\relax}        \fi
\providecommand\bibfield[2]{#2}
\providecommand\bibinfo[2]{#2}
\providecommand\natexlab[1]{#1}
\providecommand\showeprint[2][]{arXiv:#2}

\bibitem[Alwassel et~al\mbox{.}(2020)]%
        {alwassel2020self}
\bibfield{author}{\bibinfo{person}{Humam Alwassel}, \bibinfo{person}{Dhruv Mahajan}, \bibinfo{person}{Bruno Korbar}, \bibinfo{person}{Lorenzo Torresani}, \bibinfo{person}{Bernard Ghanem}, {and} \bibinfo{person}{Du Tran}.} \bibinfo{year}{2020}\natexlab{}.
\newblock \showarticletitle{Self-supervised learning by cross-modal audio-video clustering}.
\newblock \bibinfo{journal}{\emph{Advances in Neural Information Processing Systems}}  \bibinfo{volume}{33} (\bibinfo{year}{2020}), \bibinfo{pages}{9758--9770}.
\newblock


\bibitem[Andrew et~al\mbox{.}(2013)]%
        {andrew2013deep}
\bibfield{author}{\bibinfo{person}{Galen Andrew}, \bibinfo{person}{Raman Arora}, \bibinfo{person}{Jeff Bilmes}, {and} \bibinfo{person}{Karen Livescu}.} \bibinfo{year}{2013}\natexlab{}.
\newblock \showarticletitle{Deep canonical correlation analysis}. In \bibinfo{booktitle}{\emph{International conference on machine learning}}. PMLR, \bibinfo{pages}{1247--1255}.
\newblock


\bibitem[Arjovsky et~al\mbox{.}(2017)]%
        {arjovsky2017wasserstein}
\bibfield{author}{\bibinfo{person}{Martin Arjovsky}, \bibinfo{person}{Soumith Chintala}, {and} \bibinfo{person}{L{\'e}on Bottou}.} \bibinfo{year}{2017}\natexlab{}.
\newblock \showarticletitle{Wasserstein generative adversarial networks}. In \bibinfo{booktitle}{\emph{International conference on machine learning}}. PMLR, \bibinfo{pages}{214--223}.
\newblock


\bibitem[Bach and Jordan(2002)]%
        {bach2002kernel}
\bibfield{author}{\bibinfo{person}{Francis~R Bach} {and} \bibinfo{person}{Michael~I Jordan}.} \bibinfo{year}{2002}\natexlab{}.
\newblock \showarticletitle{Kernel independent component analysis}.
\newblock \bibinfo{journal}{\emph{Journal of machine learning research}} \bibinfo{volume}{3}, \bibinfo{number}{Jul} (\bibinfo{year}{2002}), \bibinfo{pages}{1--48}.
\newblock


\bibitem[Baevski et~al\mbox{.}(2020)]%
        {baevski2020wav2vec}
\bibfield{author}{\bibinfo{person}{Alexei Baevski}, \bibinfo{person}{Yuhao Zhou}, \bibinfo{person}{Abdelrahman Mohamed}, {and} \bibinfo{person}{Michael Auli}.} \bibinfo{year}{2020}\natexlab{}.
\newblock \showarticletitle{wav2vec 2.0: A framework for self-supervised learning of speech representations}.
\newblock \bibinfo{journal}{\emph{Advances in neural information processing systems}}  \bibinfo{volume}{33} (\bibinfo{year}{2020}), \bibinfo{pages}{12449--12460}.
\newblock


\bibitem[Ballan et~al\mbox{.}(2014)]%
        {ballan2014cross}
\bibfield{author}{\bibinfo{person}{Lamberto Ballan}, \bibinfo{person}{Tiberio Uricchio}, \bibinfo{person}{Lorenzo Seidenari}, {and} \bibinfo{person}{Alberto Del~Bimbo}.} \bibinfo{year}{2014}\natexlab{}.
\newblock \showarticletitle{A cross-media model for automatic image annotation}. In \bibinfo{booktitle}{\emph{Proceedings of international conference on multimedia retrieval}}. \bibinfo{pages}{73--80}.
\newblock


\bibitem[Bronstein et~al\mbox{.}(2010)]%
        {bronstein2010data}
\bibfield{author}{\bibinfo{person}{Michael~M Bronstein}, \bibinfo{person}{Alexander~M Bronstein}, \bibinfo{person}{Fabrice Michel}, {and} \bibinfo{person}{Nikos Paragios}.} \bibinfo{year}{2010}\natexlab{}.
\newblock \showarticletitle{Data fusion through cross-modality metric learning using similarity-sensitive hashing}. In \bibinfo{booktitle}{\emph{2010 IEEE computer society conference on computer vision and pattern recognition}}. IEEE, \bibinfo{pages}{3594--3601}.
\newblock


\bibitem[Chen et~al\mbox{.}(2021)]%
        {chen2021integrating}
\bibfield{author}{\bibinfo{person}{Wei Chen}, \bibinfo{person}{Yu Liu}, \bibinfo{person}{Erwin~M Bakker}, {and} \bibinfo{person}{Michael~S Lew}.} \bibinfo{year}{2021}\natexlab{}.
\newblock \showarticletitle{Integrating information theory and adversarial learning for cross-modal retrieval}.
\newblock \bibinfo{journal}{\emph{Pattern Recognition}}  \bibinfo{volume}{117} (\bibinfo{year}{2021}), \bibinfo{pages}{107983}.
\newblock


\bibitem[Cho et~al\mbox{.}(2025)]%
        {cho2025ev}
\bibfield{author}{\bibinfo{person}{Hoonhee Cho}, \bibinfo{person}{Jae-young Kang}, \bibinfo{person}{Youngho Kim}, {and} \bibinfo{person}{Kuk-Jin Yoon}.} \bibinfo{year}{2025}\natexlab{}.
\newblock \showarticletitle{Ev-3DOD: Pushing the Temporal Boundaries of 3D Object Detection with Event Cameras}.
\newblock \bibinfo{journal}{\emph{arXiv preprint arXiv:2502.19630}} (\bibinfo{year}{2025}).
\newblock


\bibitem[Chua et~al\mbox{.}(2009)]%
        {chua2009nus}
\bibfield{author}{\bibinfo{person}{Tat-Seng Chua}, \bibinfo{person}{Jinhui Tang}, \bibinfo{person}{Richang Hong}, \bibinfo{person}{Haojie Li}, \bibinfo{person}{Zhiping Luo}, {and} \bibinfo{person}{Yantao Zheng}.} \bibinfo{year}{2009}\natexlab{}.
\newblock \showarticletitle{Nus-wide: a real-world web image database from national university of singapore}. In \bibinfo{booktitle}{\emph{Proceedings of the ACM international conference on image and video retrieval}}. \bibinfo{pages}{1--9}.
\newblock


\bibitem[Clavet et~al\mbox{.}(2016)]%
        {clavet2016motion}
\bibfield{author}{\bibinfo{person}{Simon Clavet} {et~al\mbox{.}}} \bibinfo{year}{2016}\natexlab{}.
\newblock \showarticletitle{Motion matching and the road to next-gen animation}. In \bibinfo{booktitle}{\emph{Proc. of GDC}}, Vol.~\bibinfo{volume}{2}. \bibinfo{pages}{4}.
\newblock


\bibitem[De~Fauw et~al\mbox{.}(2018)]%
        {de2018clinically}
\bibfield{author}{\bibinfo{person}{Jeffrey De~Fauw}, \bibinfo{person}{Joseph~R Ledsam}, \bibinfo{person}{Bernardino Romera-Paredes}, \bibinfo{person}{Stanislav Nikolov}, \bibinfo{person}{Nenad Tomasev}, \bibinfo{person}{Sam Blackwell}, \bibinfo{person}{Harry Askham}, \bibinfo{person}{Xavier Glorot}, \bibinfo{person}{Brendan O’Donoghue}, \bibinfo{person}{Daniel Visentin}, {et~al\mbox{.}}} \bibinfo{year}{2018}\natexlab{}.
\newblock \showarticletitle{Clinically applicable deep learning for diagnosis and referral in retinal disease}.
\newblock \bibinfo{journal}{\emph{Nature medicine}} \bibinfo{volume}{24}, \bibinfo{number}{9} (\bibinfo{year}{2018}), \bibinfo{pages}{1342--1350}.
\newblock


\bibitem[Ejiga~Peter et~al\mbox{.}(2025)]%
        {ejiga2025text}
\bibfield{author}{\bibinfo{person}{Ojonugwa~Oluwafemi Ejiga~Peter}, \bibinfo{person}{Opeyemi~Taiwo Adeniran}, \bibinfo{person}{Adetokunbo~MacGregor John-Otumu}, \bibinfo{person}{Fahmi Khalifa}, {and} \bibinfo{person}{Md~Mahmudur Rahman}.} \bibinfo{year}{2025}\natexlab{}.
\newblock \showarticletitle{Text-Guided Synthesis in Medical Multimedia Retrieval: A Framework for Enhanced Colonoscopy Image Classification and Segmentation}.
\newblock \bibinfo{journal}{\emph{Algorithms}} \bibinfo{volume}{18}, \bibinfo{number}{3} (\bibinfo{year}{2025}), \bibinfo{pages}{155}.
\newblock


\bibitem[Felix et~al\mbox{.}(2018)]%
        {felix2018multi}
\bibfield{author}{\bibinfo{person}{Rafael Felix}, \bibinfo{person}{Ian Reid}, \bibinfo{person}{Gustavo Carneiro}, {et~al\mbox{.}}} \bibinfo{year}{2018}\natexlab{}.
\newblock \showarticletitle{Multi-modal cycle-consistent generalized zero-shot learning}. In \bibinfo{booktitle}{\emph{Proceedings of the European conference on computer vision (ECCV)}}. \bibinfo{pages}{21--37}.
\newblock


\bibitem[Feng et~al\mbox{.}(2023)]%
        {feng2023rono}
\bibfield{author}{\bibinfo{person}{Yanglin Feng}, \bibinfo{person}{Hongyuan Zhu}, \bibinfo{person}{Dezhong Peng}, \bibinfo{person}{Xi Peng}, {and} \bibinfo{person}{Peng Hu}.} \bibinfo{year}{2023}\natexlab{}.
\newblock \showarticletitle{RONO: robust discriminative learning with noisy labels for 2D-3D cross-modal retrieval}. In \bibinfo{booktitle}{\emph{Proceedings of the IEEE/CVF Conference on Computer Vision and Pattern Recognition}}. \bibinfo{pages}{11610--11619}.
\newblock


\bibitem[Finner(1992)]%
        {finner1992generalization}
\bibfield{author}{\bibinfo{person}{Helmut Finner}.} \bibinfo{year}{1992}\natexlab{}.
\newblock \showarticletitle{A generalization of H{\"o}lder's inequality and some probability inequalities}.
\newblock \bibinfo{journal}{\emph{The Annals of probability}} (\bibinfo{year}{1992}), \bibinfo{pages}{1893--1901}.
\newblock


\bibitem[Goodfellow et~al\mbox{.}(2014)]%
        {goodfellow2014generative}
\bibfield{author}{\bibinfo{person}{Ian~J Goodfellow}, \bibinfo{person}{Jean Pouget-Abadie}, \bibinfo{person}{Mehdi Mirza}, \bibinfo{person}{Bing Xu}, \bibinfo{person}{David Warde-Farley}, \bibinfo{person}{Sherjil Ozair}, \bibinfo{person}{Aaron Courville}, {and} \bibinfo{person}{Yoshua Bengio}.} \bibinfo{year}{2014}\natexlab{}.
\newblock \showarticletitle{Generative adversarial nets}.
\newblock \bibinfo{journal}{\emph{Advances in neural information processing systems}}  \bibinfo{volume}{27} (\bibinfo{year}{2014}).
\newblock


\bibitem[Gretton et~al\mbox{.}(2006)]%
        {gretton2006kernel}
\bibfield{author}{\bibinfo{person}{Arthur Gretton}, \bibinfo{person}{Karsten Borgwardt}, \bibinfo{person}{Malte Rasch}, \bibinfo{person}{Bernhard Sch{\"o}lkopf}, {and} \bibinfo{person}{Alex Smola}.} \bibinfo{year}{2006}\natexlab{}.
\newblock \showarticletitle{A kernel method for the two-sample-problem}.
\newblock \bibinfo{journal}{\emph{Advances in neural information processing systems}}  \bibinfo{volume}{19} (\bibinfo{year}{2006}).
\newblock


\bibitem[He et~al\mbox{.}(2022)]%
        {he2022category}
\bibfield{author}{\bibinfo{person}{Shiyuan He}, \bibinfo{person}{Weiyang Wang}, \bibinfo{person}{Zheng Wang}, \bibinfo{person}{Xing Xu}, \bibinfo{person}{Yang Yang}, \bibinfo{person}{Xiaoming Wang}, {and} \bibinfo{person}{Heng~Tao Shen}.} \bibinfo{year}{2022}\natexlab{}.
\newblock \showarticletitle{Category alignment adversarial learning for cross-modal retrieval}.
\newblock \bibinfo{journal}{\emph{IEEE Transactions on Knowledge and Data Engineering}} \bibinfo{volume}{35}, \bibinfo{number}{5} (\bibinfo{year}{2022}), \bibinfo{pages}{4527--4538}.
\newblock


\bibitem[Hotelling(1992)]%
        {hotelling1992relations}
\bibfield{author}{\bibinfo{person}{Harold Hotelling}.} \bibinfo{year}{1992}\natexlab{}.
\newblock \showarticletitle{Relations between two sets of variates}.
\newblock In \bibinfo{booktitle}{\emph{Breakthroughs in statistics: methodology and distribution}}. \bibinfo{publisher}{Springer}, \bibinfo{pages}{162--190}.
\newblock


\bibitem[Huang et~al\mbox{.}(2018)]%
        {huang2018mhtn}
\bibfield{author}{\bibinfo{person}{Xin Huang}, \bibinfo{person}{Yuxin Peng}, {and} \bibinfo{person}{Mingkuan Yuan}.} \bibinfo{year}{2018}\natexlab{}.
\newblock \showarticletitle{MHTN: Modal-adversarial hybrid transfer network for cross-modal retrieval}.
\newblock \bibinfo{journal}{\emph{IEEE transactions on cybernetics}} \bibinfo{volume}{50}, \bibinfo{number}{3} (\bibinfo{year}{2018}), \bibinfo{pages}{1047--1059}.
\newblock


\bibitem[Jambigi et~al\mbox{.}(2021)]%
        {jambigi2021mmd}
\bibfield{author}{\bibinfo{person}{Chaitra Jambigi}, \bibinfo{person}{Ruchit Rawal}, {and} \bibinfo{person}{Anirban Chakraborty}.} \bibinfo{year}{2021}\natexlab{}.
\newblock \showarticletitle{Mmd-reid: A simple but effective solution for visible-thermal person reid}.
\newblock \bibinfo{journal}{\emph{arXiv preprint arXiv:2111.05059}} (\bibinfo{year}{2021}).
\newblock


\bibitem[Jenssen et~al\mbox{.}(2006)]%
        {jenssen2006cauchy}
\bibfield{author}{\bibinfo{person}{Robert Jenssen}, \bibinfo{person}{Jose~C Principe}, \bibinfo{person}{Deniz Erdogmus}, {and} \bibinfo{person}{Torbj{\o}rn Eltoft}.} \bibinfo{year}{2006}\natexlab{}.
\newblock \showarticletitle{The Cauchy--Schwarz divergence and Parzen windowing: Connections to graph theory and Mercer kernels}.
\newblock \bibinfo{journal}{\emph{Journal of the Franklin Institute}} \bibinfo{volume}{343}, \bibinfo{number}{6} (\bibinfo{year}{2006}), \bibinfo{pages}{614--629}.
\newblock


\bibitem[Jia et~al\mbox{.}(2019)]%
        {jia2019semantically}
\bibfield{author}{\bibinfo{person}{Yuhua Jia}, \bibinfo{person}{Liang Bai}, \bibinfo{person}{Shuang Liu}, \bibinfo{person}{Peng Wang}, \bibinfo{person}{Jinlin Guo}, {and} \bibinfo{person}{Yuxiang Xie}.} \bibinfo{year}{2019}\natexlab{}.
\newblock \showarticletitle{Semantically-enhanced kernel canonical correlation analysis: a multi-label cross-modal retrieval}.
\newblock \bibinfo{journal}{\emph{Multimedia Tools and Applications}}  \bibinfo{volume}{78} (\bibinfo{year}{2019}), \bibinfo{pages}{13169--13188}.
\newblock


\bibitem[Jiang and Ye(2023)]%
        {jiang2023cross}
\bibfield{author}{\bibinfo{person}{Ding Jiang} {and} \bibinfo{person}{Mang Ye}.} \bibinfo{year}{2023}\natexlab{}.
\newblock \showarticletitle{Cross-modal implicit relation reasoning and aligning for text-to-image person retrieval}. In \bibinfo{booktitle}{\emph{Proceedings of the IEEE/CVF Conference on Computer Vision and Pattern Recognition}}. \bibinfo{pages}{2787--2797}.
\newblock


\bibitem[Jiang and Li(2017)]%
        {jiang2017deep}
\bibfield{author}{\bibinfo{person}{Qing-Yuan Jiang} {and} \bibinfo{person}{Wu-Jun Li}.} \bibinfo{year}{2017}\natexlab{}.
\newblock \showarticletitle{Deep cross-modal hashing}. In \bibinfo{booktitle}{\emph{Proceedings of the IEEE conference on computer vision and pattern recognition}}. \bibinfo{pages}{3232--3240}.
\newblock


\bibitem[Kampa et~al\mbox{.}(2011)]%
        {kampa2011closed}
\bibfield{author}{\bibinfo{person}{Kittipat Kampa}, \bibinfo{person}{Erion Hasanbelliu}, {and} \bibinfo{person}{Jose~C Principe}.} \bibinfo{year}{2011}\natexlab{}.
\newblock \showarticletitle{Closed-form Cauchy-Schwarz PDF divergence for mixture of Gaussians}. In \bibinfo{booktitle}{\emph{The 2011 International Joint Conference on Neural Networks}}. IEEE, \bibinfo{pages}{2578--2585}.
\newblock


\bibitem[Khan et~al\mbox{.}(2024)]%
        {khan2024invideo}
\bibfield{author}{\bibinfo{person}{Almira~Asif Khan}, \bibinfo{person}{Muhammed}, \bibinfo{person}{Asher~Mathews Shaji}, \bibinfo{person}{Devika Sujith}, \bibinfo{person}{Aneesh~G Nath}, {and} \bibinfo{person}{Sandeep~S Udmale}.} \bibinfo{year}{2024}\natexlab{}.
\newblock \showarticletitle{InVideo Search: Scene Description Clustering and Integrating Image and Audio Captioning for Enhanced Video Search}. In \bibinfo{booktitle}{\emph{International Conference on Distributed Computing and Intelligent Technology}}. Springer, \bibinfo{pages}{195--208}.
\newblock


\bibitem[Kim et~al\mbox{.}(2023)]%
        {kim2023improving}
\bibfield{author}{\bibinfo{person}{Dongwon Kim}, \bibinfo{person}{Namyup Kim}, {and} \bibinfo{person}{Suha Kwak}.} \bibinfo{year}{2023}\natexlab{}.
\newblock \showarticletitle{Improving cross-modal retrieval with set of diverse embeddings}. In \bibinfo{booktitle}{\emph{Proceedings of the IEEE/CVF conference on computer vision and pattern recognition}}. \bibinfo{pages}{23422--23431}.
\newblock


\bibitem[Kullback and Leibler(1951)]%
        {kullback1951information}
\bibfield{author}{\bibinfo{person}{Solomon Kullback} {and} \bibinfo{person}{Richard~A Leibler}.} \bibinfo{year}{1951}\natexlab{}.
\newblock \showarticletitle{On information and sufficiency}.
\newblock \bibinfo{journal}{\emph{The annals of mathematical statistics}} \bibinfo{volume}{22}, \bibinfo{number}{1} (\bibinfo{year}{1951}), \bibinfo{pages}{79--86}.
\newblock


\bibitem[Li et~al\mbox{.}(2003)]%
        {li2003multimedia}
\bibfield{author}{\bibinfo{person}{Dongge Li}, \bibinfo{person}{Nevenka Dimitrova}, \bibinfo{person}{Mingkun Li}, {and} \bibinfo{person}{Ishwar~K Sethi}.} \bibinfo{year}{2003}\natexlab{}.
\newblock \showarticletitle{Multimedia content processing through cross-modal association}. In \bibinfo{booktitle}{\emph{Proceedings of the eleventh ACM international conference on Multimedia}}. \bibinfo{pages}{604--611}.
\newblock


\bibitem[Li et~al\mbox{.}(2017)]%
        {li2017person}
\bibfield{author}{\bibinfo{person}{Shuang Li}, \bibinfo{person}{Tong Xiao}, \bibinfo{person}{Hongsheng Li}, \bibinfo{person}{Bolei Zhou}, \bibinfo{person}{Dayu Yue}, {and} \bibinfo{person}{Xiaogang Wang}.} \bibinfo{year}{2017}\natexlab{}.
\newblock \showarticletitle{Person search with natural language description}. In \bibinfo{booktitle}{\emph{Proceedings of the IEEE conference on computer vision and pattern recognition}}. \bibinfo{pages}{1970--1979}.
\newblock


\bibitem[Li et~al\mbox{.}(2015)]%
        {li2015generative}
\bibfield{author}{\bibinfo{person}{Yujia Li}, \bibinfo{person}{Kevin Swersky}, {and} \bibinfo{person}{Rich Zemel}.} \bibinfo{year}{2015}\natexlab{}.
\newblock \showarticletitle{Generative moment matching networks}. In \bibinfo{booktitle}{\emph{International conference on machine learning}}. PMLR, \bibinfo{pages}{1718--1727}.
\newblock


\bibitem[Li et~al\mbox{.}(2024)]%
        {li2024supervised}
\bibfield{author}{\bibinfo{person}{Ze Li}, \bibinfo{person}{Tao Yao}, \bibinfo{person}{Lili Wang}, \bibinfo{person}{Ying Li}, {and} \bibinfo{person}{Gang Wang}.} \bibinfo{year}{2024}\natexlab{}.
\newblock \showarticletitle{Supervised Contrastive Discrete Hashing for cross-modal retrieval}.
\newblock \bibinfo{journal}{\emph{Knowledge-Based Systems}}  \bibinfo{volume}{295} (\bibinfo{year}{2024}), \bibinfo{pages}{111837}.
\newblock


\bibitem[Liang et~al\mbox{.}(2022)]%
        {liang2022mind}
\bibfield{author}{\bibinfo{person}{Victor~Weixin Liang}, \bibinfo{person}{Yuhui Zhang}, \bibinfo{person}{Yongchan Kwon}, \bibinfo{person}{Serena Yeung}, {and} \bibinfo{person}{James~Y Zou}.} \bibinfo{year}{2022}\natexlab{}.
\newblock \showarticletitle{Mind the gap: Understanding the modality gap in multi-modal contrastive representation learning}.
\newblock \bibinfo{journal}{\emph{Advances in Neural Information Processing Systems}}  \bibinfo{volume}{35} (\bibinfo{year}{2022}), \bibinfo{pages}{17612--17625}.
\newblock


\bibitem[Lin et~al\mbox{.}(2015)]%
        {lin2015semantics}
\bibfield{author}{\bibinfo{person}{Zijia Lin}, \bibinfo{person}{Guiguang Ding}, \bibinfo{person}{Mingqing Hu}, {and} \bibinfo{person}{Jianmin Wang}.} \bibinfo{year}{2015}\natexlab{}.
\newblock \showarticletitle{Semantics-preserving hashing for cross-view retrieval}. In \bibinfo{booktitle}{\emph{Proceedings of the IEEE conference on computer vision and pattern recognition}}. \bibinfo{pages}{3864--3872}.
\newblock


\bibitem[Liu et~al\mbox{.}(2018)]%
        {liu2018index}
\bibfield{author}{\bibinfo{person}{Luchen Liu}, \bibinfo{person}{Yang Yang}, \bibinfo{person}{Mengqiu Hu}, \bibinfo{person}{Xing Xu}, \bibinfo{person}{Fumin Shen}, \bibinfo{person}{Ning Xie}, {and} \bibinfo{person}{Zi Huang}.} \bibinfo{year}{2018}\natexlab{}.
\newblock \showarticletitle{Index and retrieve multimedia data: Cross-modal hashing by learning subspace relation}. In \bibinfo{booktitle}{\emph{Database Systems for Advanced Applications: 23rd International Conference, DASFAA 2018, Gold Coast, QLD, Australia, May 21-24, 2018, Proceedings, Part II 23}}. Springer, \bibinfo{pages}{606--621}.
\newblock


\bibitem[Liu et~al\mbox{.}(2024)]%
        {liu2024multimodal}
\bibfield{author}{\bibinfo{person}{Qidong Liu}, \bibinfo{person}{Jiaxi Hu}, \bibinfo{person}{Yutian Xiao}, \bibinfo{person}{Xiangyu Zhao}, \bibinfo{person}{Jingtong Gao}, \bibinfo{person}{Wanyu Wang}, \bibinfo{person}{Qing Li}, {and} \bibinfo{person}{Jiliang Tang}.} \bibinfo{year}{2024}\natexlab{}.
\newblock \showarticletitle{Multimodal recommender systems: A survey}.
\newblock \bibinfo{journal}{\emph{Comput. Surveys}} \bibinfo{volume}{57}, \bibinfo{number}{2} (\bibinfo{year}{2024}), \bibinfo{pages}{1--17}.
\newblock


\bibitem[Mikolov et~al\mbox{.}(2013)]%
        {mikolov2013efficient}
\bibfield{author}{\bibinfo{person}{Tomas Mikolov}, \bibinfo{person}{Kai Chen}, \bibinfo{person}{Greg Corrado}, {and} \bibinfo{person}{Jeffrey Dean}.} \bibinfo{year}{2013}\natexlab{}.
\newblock \showarticletitle{Efficient estimation of word representations in vector space}.
\newblock \bibinfo{journal}{\emph{arXiv preprint arXiv:1301.3781}} (\bibinfo{year}{2013}).
\newblock


\bibitem[Peng and Qi(2019)]%
        {peng2019cm}
\bibfield{author}{\bibinfo{person}{Yuxin Peng} {and} \bibinfo{person}{Jinwei Qi}.} \bibinfo{year}{2019}\natexlab{}.
\newblock \showarticletitle{CM-GANs: Cross-modal generative adversarial networks for common representation learning}.
\newblock \bibinfo{journal}{\emph{ACM Transactions on Multimedia Computing, Communications, and Applications (TOMM)}} \bibinfo{volume}{15}, \bibinfo{number}{1} (\bibinfo{year}{2019}), \bibinfo{pages}{1--24}.
\newblock


\bibitem[Peng et~al\mbox{.}(2017)]%
        {peng2017ccl}
\bibfield{author}{\bibinfo{person}{Yuxin Peng}, \bibinfo{person}{Jinwei Qi}, \bibinfo{person}{Xin Huang}, {and} \bibinfo{person}{Yuxin Yuan}.} \bibinfo{year}{2017}\natexlab{}.
\newblock \showarticletitle{CCL: Cross-modal correlation learning with multigrained fusion by hierarchical network}.
\newblock \bibinfo{journal}{\emph{IEEE Transactions on Multimedia}} \bibinfo{volume}{20}, \bibinfo{number}{2} (\bibinfo{year}{2017}), \bibinfo{pages}{405--420}.
\newblock


\bibitem[Peng et~al\mbox{.}(2020)]%
        {peng2020unsupervised}
\bibfield{author}{\bibinfo{person}{Yuxin Peng}, \bibinfo{person}{Zhaoda Ye}, \bibinfo{person}{Jinwei Qi}, {and} \bibinfo{person}{Yunkan Zhuo}.} \bibinfo{year}{2020}\natexlab{}.
\newblock \showarticletitle{Unsupervised visual--textual correlation learning with fine-grained semantic alignment}.
\newblock \bibinfo{journal}{\emph{IEEE Transactions on Cybernetics}} \bibinfo{volume}{52}, \bibinfo{number}{5} (\bibinfo{year}{2020}), \bibinfo{pages}{3669--3683}.
\newblock


\bibitem[Petrovich et~al\mbox{.}(2023)]%
        {petrovich2023tmr}
\bibfield{author}{\bibinfo{person}{Mathis Petrovich}, \bibinfo{person}{Michael~J Black}, {and} \bibinfo{person}{G{\"u}l Varol}.} \bibinfo{year}{2023}\natexlab{}.
\newblock \showarticletitle{Tmr: Text-to-motion retrieval using contrastive 3d human motion synthesis}. In \bibinfo{booktitle}{\emph{Proceedings of the IEEE/CVF International Conference on Computer Vision}}. \bibinfo{pages}{9488--9497}.
\newblock


\bibitem[Plappert et~al\mbox{.}(2016)]%
        {plappert2016kit}
\bibfield{author}{\bibinfo{person}{Matthias Plappert}, \bibinfo{person}{Christian Mandery}, {and} \bibinfo{person}{Tamim Asfour}.} \bibinfo{year}{2016}\natexlab{}.
\newblock \showarticletitle{The kit motion-language dataset}.
\newblock \bibinfo{journal}{\emph{Big data}} \bibinfo{volume}{4}, \bibinfo{number}{4} (\bibinfo{year}{2016}), \bibinfo{pages}{236--252}.
\newblock


\bibitem[Principe et~al\mbox{.}(2000)]%
        {principe2000information}
\bibfield{author}{\bibinfo{person}{Jose~C Principe}, \bibinfo{person}{Dongxin Xu}, \bibinfo{person}{J Fisher}, {and} \bibinfo{person}{S Haykin}.} \bibinfo{year}{2000}\natexlab{}.
\newblock \showarticletitle{Information theoretic learning. Unsupervised adaptive filtering}.
\newblock \bibinfo{journal}{\emph{Unsupervised Adapt Filter}}  \bibinfo{volume}{1} (\bibinfo{year}{2000}).
\newblock


\bibitem[Pu et~al\mbox{.}(2025)]%
        {pu2025deep}
\bibfield{author}{\bibinfo{person}{Ruitao Pu}, \bibinfo{person}{Yang Qin}, \bibinfo{person}{Dezhong Peng}, \bibinfo{person}{Xiaomin Song}, {and} \bibinfo{person}{Huiming Zheng}.} \bibinfo{year}{2025}\natexlab{}.
\newblock \showarticletitle{Deep reversible consistency learning for cross-modal retrieval}.
\newblock \bibinfo{journal}{\emph{IEEE Transactions on Multimedia}} (\bibinfo{year}{2025}).
\newblock


\bibitem[Qiang and Hu(2011)]%
        {qiang2011generalizations}
\bibfield{author}{\bibinfo{person}{Hua Qiang} {and} \bibinfo{person}{Zhicheng Hu}.} \bibinfo{year}{2011}\natexlab{}.
\newblock \showarticletitle{Generalizations of H{\"o}lder’s and some related inequalities}.
\newblock \bibinfo{journal}{\emph{Computers \& Mathematics with Applications}} \bibinfo{volume}{61}, \bibinfo{number}{2} (\bibinfo{year}{2011}), \bibinfo{pages}{392--396}.
\newblock


\bibitem[Quan et~al\mbox{.}(2025)]%
        {quan2025vifi}
\bibfield{author}{\bibinfo{person}{Khanh-An~C Quan}, \bibinfo{person}{Qui~Ngoc Nguyen}, {and} \bibinfo{person}{Minh-Triet Tran}.} \bibinfo{year}{2025}\natexlab{}.
\newblock \showarticletitle{ViFi: A Video Finding System at Video Browser Showdown 2025}. In \bibinfo{booktitle}{\emph{International Conference on Multimedia Modeling}}. Springer, \bibinfo{pages}{378--384}.
\newblock


\bibitem[Radford et~al\mbox{.}(2021)]%
        {radford2021learning}
\bibfield{author}{\bibinfo{person}{Alec Radford}, \bibinfo{person}{Jong~Wook Kim}, \bibinfo{person}{Chris Hallacy}, \bibinfo{person}{Aditya Ramesh}, \bibinfo{person}{Gabriel Goh}, \bibinfo{person}{Sandhini Agarwal}, \bibinfo{person}{Girish Sastry}, \bibinfo{person}{Amanda Askell}, \bibinfo{person}{Pamela Mishkin}, \bibinfo{person}{Jack Clark}, {et~al\mbox{.}}} \bibinfo{year}{2021}\natexlab{}.
\newblock \showarticletitle{Learning transferable visual models from natural language supervision}. In \bibinfo{booktitle}{\emph{International conference on machine learning}}. PMLR, \bibinfo{pages}{8748--8763}.
\newblock


\bibitem[Ramesh et~al\mbox{.}(2021)]%
        {ramesh2021zero}
\bibfield{author}{\bibinfo{person}{Aditya Ramesh}, \bibinfo{person}{Mikhail Pavlov}, \bibinfo{person}{Gabriel Goh}, \bibinfo{person}{Scott Gray}, \bibinfo{person}{Chelsea Voss}, \bibinfo{person}{Alec Radford}, \bibinfo{person}{Mark Chen}, {and} \bibinfo{person}{Ilya Sutskever}.} \bibinfo{year}{2021}\natexlab{}.
\newblock \showarticletitle{Zero-shot text-to-image generation}. In \bibinfo{booktitle}{\emph{International conference on machine learning}}. Pmlr, \bibinfo{pages}{8821--8831}.
\newblock


\bibitem[Rasiwasia et~al\mbox{.}(2010)]%
        {rasiwasia2010new}
\bibfield{author}{\bibinfo{person}{Nikhil Rasiwasia}, \bibinfo{person}{Jose Costa~Pereira}, \bibinfo{person}{Emanuele Coviello}, \bibinfo{person}{Gabriel Doyle}, \bibinfo{person}{Gert~RG Lanckriet}, \bibinfo{person}{Roger Levy}, {and} \bibinfo{person}{Nuno Vasconcelos}.} \bibinfo{year}{2010}\natexlab{}.
\newblock \showarticletitle{A new approach to cross-modal multimedia retrieval}. In \bibinfo{booktitle}{\emph{Proceedings of the 18th ACM international conference on Multimedia}}. \bibinfo{pages}{251--260}.
\newblock


\bibitem[Salehi et~al\mbox{.}(2024)]%
        {salehi2024flash}
\bibfield{author}{\bibinfo{person}{Batool Salehi}, \bibinfo{person}{Debashri Roy}, \bibinfo{person}{Jerry Gu}, \bibinfo{person}{Chris Dick}, {and} \bibinfo{person}{Kaushik Chowdhury}.} \bibinfo{year}{2024}\natexlab{}.
\newblock \showarticletitle{Flash-and-prune: Federated learning for automated selection of high-band mmwave sectors using model pruning}.
\newblock \bibinfo{journal}{\emph{IEEE Transactions on Mobile Computing}} (\bibinfo{year}{2024}).
\newblock


\bibitem[Sanh et~al\mbox{.}(2019)]%
        {sanh2019distilbert}
\bibfield{author}{\bibinfo{person}{Victor Sanh}, \bibinfo{person}{Lysandre Debut}, \bibinfo{person}{Julien Chaumond}, {and} \bibinfo{person}{Thomas Wolf}.} \bibinfo{year}{2019}\natexlab{}.
\newblock \showarticletitle{DistilBERT, a distilled version of BERT: smaller, faster, cheaper and lighter}.
\newblock \bibinfo{journal}{\emph{arXiv preprint arXiv:1910.01108}} (\bibinfo{year}{2019}).
\newblock


\bibitem[Shi et~al\mbox{.}(2022)]%
        {shi2022information}
\bibfield{author}{\bibinfo{person}{Yufeng Shi}, \bibinfo{person}{Shujian Yu}, \bibinfo{person}{Duanquan Xu}, {and} \bibinfo{person}{Xinge You}.} \bibinfo{year}{2022}\natexlab{}.
\newblock \showarticletitle{Information-Theoretic Hashing for Zero-Shot Cross-Modal Retrieval}.
\newblock \bibinfo{journal}{\emph{arXiv preprint arXiv:2209.12491}} (\bibinfo{year}{2022}).
\newblock


\bibitem[Shu and Zhao(2021)]%
        {shu2021scalable}
\bibfield{author}{\bibinfo{person}{Xin Shu} {and} \bibinfo{person}{Guoying Zhao}.} \bibinfo{year}{2021}\natexlab{}.
\newblock \showarticletitle{Scalable multi-label canonical correlation analysis for cross-modal retrieval}.
\newblock \bibinfo{journal}{\emph{Pattern Recognition}}  \bibinfo{volume}{115} (\bibinfo{year}{2021}), \bibinfo{pages}{107905}.
\newblock


\bibitem[Simonyan and Zisserman(2014)]%
        {simonyan2014very}
\bibfield{author}{\bibinfo{person}{Karen Simonyan} {and} \bibinfo{person}{Andrew Zisserman}.} \bibinfo{year}{2014}\natexlab{}.
\newblock \showarticletitle{Very deep convolutional networks for large-scale image recognition}.
\newblock \bibinfo{journal}{\emph{arXiv preprint arXiv:1409.1556}} (\bibinfo{year}{2014}).
\newblock


\bibitem[Sun et~al\mbox{.}(2016)]%
        {sun2016return}
\bibfield{author}{\bibinfo{person}{Baochen Sun}, \bibinfo{person}{Jiashi Feng}, {and} \bibinfo{person}{Kate Saenko}.} \bibinfo{year}{2016}\natexlab{}.
\newblock \showarticletitle{Return of frustratingly easy domain adaptation}. In \bibinfo{booktitle}{\emph{Proceedings of the AAAI conference on artificial intelligence}}, Vol.~\bibinfo{volume}{30}.
\newblock


\bibitem[Tevet et~al\mbox{.}(2022)]%
        {tevet2022motionclip}
\bibfield{author}{\bibinfo{person}{Guy Tevet}, \bibinfo{person}{Brian Gordon}, \bibinfo{person}{Amir Hertz}, \bibinfo{person}{Amit~H Bermano}, {and} \bibinfo{person}{Daniel Cohen-Or}.} \bibinfo{year}{2022}\natexlab{}.
\newblock \showarticletitle{Motionclip: Exposing human motion generation to clip space}. In \bibinfo{booktitle}{\emph{European Conference on Computer Vision}}. Springer, \bibinfo{pages}{358--374}.
\newblock


\bibitem[Trosten et~al\mbox{.}(2021)]%
        {trosten2021reconsidering}
\bibfield{author}{\bibinfo{person}{Daniel~J Trosten}, \bibinfo{person}{Sigurd Lokse}, \bibinfo{person}{Robert Jenssen}, {and} \bibinfo{person}{Michael Kampffmeyer}.} \bibinfo{year}{2021}\natexlab{}.
\newblock \showarticletitle{Reconsidering representation alignment for multi-view clustering}. In \bibinfo{booktitle}{\emph{Proceedings of the IEEE/CVF conference on computer vision and pattern recognition}}. \bibinfo{pages}{1255--1265}.
\newblock


\bibitem[Van~der Maaten and Hinton(2008)]%
        {van2008visualizing}
\bibfield{author}{\bibinfo{person}{Laurens Van~der Maaten} {and} \bibinfo{person}{Geoffrey Hinton}.} \bibinfo{year}{2008}\natexlab{}.
\newblock \showarticletitle{Visualizing data using t-SNE.}
\newblock \bibinfo{journal}{\emph{Journal of machine learning research}} \bibinfo{volume}{9}, \bibinfo{number}{11} (\bibinfo{year}{2008}).
\newblock


\bibitem[Wang et~al\mbox{.}(2017)]%
        {wang2017adversarial}
\bibfield{author}{\bibinfo{person}{Bokun Wang}, \bibinfo{person}{Yang Yang}, \bibinfo{person}{Xing Xu}, \bibinfo{person}{Alan Hanjalic}, {and} \bibinfo{person}{Heng~Tao Shen}.} \bibinfo{year}{2017}\natexlab{}.
\newblock \showarticletitle{Adversarial cross-modal retrieval}. In \bibinfo{booktitle}{\emph{Proceedings of the 25th ACM international conference on Multimedia}}. \bibinfo{pages}{154--162}.
\newblock


\bibitem[Wang et~al\mbox{.}(2025)]%
        {wang2025cross}
\bibfield{author}{\bibinfo{person}{Tianshi Wang}, \bibinfo{person}{Fengling Li}, \bibinfo{person}{Lei Zhu}, \bibinfo{person}{Jingjing Li}, \bibinfo{person}{Zheng Zhang}, {and} \bibinfo{person}{Heng~Tao Shen}.} \bibinfo{year}{2025}\natexlab{}.
\newblock \showarticletitle{Cross-modal retrieval: a systematic review of methods and future directions}.
\newblock \bibinfo{journal}{\emph{Proc. IEEE}} (\bibinfo{year}{2025}).
\newblock


\bibitem[Wang et~al\mbox{.}(2015)]%
        {wang2015unsupervised}
\bibfield{author}{\bibinfo{person}{Weiran Wang}, \bibinfo{person}{Raman Arora}, \bibinfo{person}{Karen Livescu}, {and} \bibinfo{person}{Jeff~A Bilmes}.} \bibinfo{year}{2015}\natexlab{}.
\newblock \showarticletitle{Unsupervised learning of acoustic features via deep canonical correlation analysis}. In \bibinfo{booktitle}{\emph{2015 IEEE International Conference on Acoustics, Speech and Signal Processing (ICASSP)}}. IEEE, \bibinfo{pages}{4590--4594}.
\newblock


\bibitem[Wei et~al\mbox{.}(2024)]%
        {wei2024cross}
\bibfield{author}{\bibinfo{person}{Yuxin Wei}, \bibinfo{person}{Ligang Zheng}, \bibinfo{person}{Guoping Qiu}, {and} \bibinfo{person}{Guocan Cai}.} \bibinfo{year}{2024}\natexlab{}.
\newblock \showarticletitle{Cross-modal retrieval based on shared proxies}.
\newblock \bibinfo{journal}{\emph{International Journal of Multimedia Information Retrieval}} \bibinfo{volume}{13}, \bibinfo{number}{1} (\bibinfo{year}{2024}), \bibinfo{pages}{5}.
\newblock


\bibitem[Wojcik et~al\mbox{.}(2024)]%
        {wojcik2024case}
\bibfield{author}{\bibinfo{person}{Jagoda Wojcik}, \bibinfo{person}{Jiaqi Jiang}, \bibinfo{person}{Jiacheng Wu}, {and} \bibinfo{person}{Shan Luo}.} \bibinfo{year}{2024}\natexlab{}.
\newblock \showarticletitle{A Case Study on Visual-Audio-Tactile Cross-Modal Retrieval}. In \bibinfo{booktitle}{\emph{2024 IEEE/RSJ International Conference on Intelligent Robots and Systems (IROS)}}. IEEE, \bibinfo{pages}{12472--12478}.
\newblock


\bibitem[Wu et~al\mbox{.}(2023)]%
        {wu2023large}
\bibfield{author}{\bibinfo{person}{Yusong Wu}, \bibinfo{person}{Ke Chen}, \bibinfo{person}{Tianyu Zhang}, \bibinfo{person}{Yuchen Hui}, \bibinfo{person}{Taylor Berg-Kirkpatrick}, {and} \bibinfo{person}{Shlomo Dubnov}.} \bibinfo{year}{2023}\natexlab{}.
\newblock \showarticletitle{Large-scale contrastive language-audio pretraining with feature fusion and keyword-to-caption augmentation}. In \bibinfo{booktitle}{\emph{ICASSP 2023-2023 IEEE International Conference on Acoustics, Speech and Signal Processing (ICASSP)}}. IEEE, \bibinfo{pages}{1--5}.
\newblock


\bibitem[Xi et~al\mbox{.}(2025)]%
        {xi2025cross}
\bibfield{author}{\bibinfo{person}{Xiaocong Xi}, \bibinfo{person}{Chee-Onn Chow}, \bibinfo{person}{Joon~Huang Chuah}, {and} \bibinfo{person}{Jeevan Kanesan}.} \bibinfo{year}{2025}\natexlab{}.
\newblock \showarticletitle{Cross-Modal Semantic Relations Enhancement with Graph Attention Network for Image-Text Matching}.
\newblock \bibinfo{journal}{\emph{IEEE Access}} \bibinfo{number}{99} (\bibinfo{year}{2025}), \bibinfo{pages}{1--1}.
\newblock


\bibitem[Xu et~al\mbox{.}(2020)]%
        {xu2020joint}
\bibfield{author}{\bibinfo{person}{Xing Xu}, \bibinfo{person}{Kaiyi Lin}, \bibinfo{person}{Yang Yang}, \bibinfo{person}{Alan Hanjalic}, {and} \bibinfo{person}{Heng~Tao Shen}.} \bibinfo{year}{2020}\natexlab{}.
\newblock \showarticletitle{Joint feature synthesis and embedding: Adversarial cross-modal retrieval revisited}.
\newblock \bibinfo{journal}{\emph{IEEE Transactions on Pattern Analysis and Machine Intelligence}} \bibinfo{volume}{44}, \bibinfo{number}{6} (\bibinfo{year}{2020}), \bibinfo{pages}{3030--3047}.
\newblock


\bibitem[Xu et~al\mbox{.}(2019)]%
        {xu2019ternary}
\bibfield{author}{\bibinfo{person}{Xing Xu}, \bibinfo{person}{Huimin Lu}, \bibinfo{person}{Jingkuan Song}, \bibinfo{person}{Yang Yang}, \bibinfo{person}{Heng~Tao Shen}, {and} \bibinfo{person}{Xuelong Li}.} \bibinfo{year}{2019}\natexlab{}.
\newblock \showarticletitle{Ternary adversarial networks with self-supervision for zero-shot cross-modal retrieval}.
\newblock \bibinfo{journal}{\emph{IEEE transactions on cybernetics}} \bibinfo{volume}{50}, \bibinfo{number}{6} (\bibinfo{year}{2019}), \bibinfo{pages}{2400--2413}.
\newblock


\bibitem[Yan and Mikolajczyk(2015)]%
        {yan2015deep}
\bibfield{author}{\bibinfo{person}{Fei Yan} {and} \bibinfo{person}{Krystian Mikolajczyk}.} \bibinfo{year}{2015}\natexlab{}.
\newblock \showarticletitle{Deep correlation for matching images and text}. In \bibinfo{booktitle}{\emph{Proceedings of the IEEE conference on computer vision and pattern recognition}}. \bibinfo{pages}{3441--3450}.
\newblock


\bibitem[Yang et~al\mbox{.}(2025)]%
        {yang2025egolife}
\bibfield{author}{\bibinfo{person}{Jingkang Yang}, \bibinfo{person}{Shuai Liu}, \bibinfo{person}{Hongming Guo}, \bibinfo{person}{Yuhao Dong}, \bibinfo{person}{Xiamengwei Zhang}, \bibinfo{person}{Sicheng Zhang}, \bibinfo{person}{Pengyun Wang}, \bibinfo{person}{Zitang Zhou}, \bibinfo{person}{Binzhu Xie}, \bibinfo{person}{Ziyue Wang}, {et~al\mbox{.}}} \bibinfo{year}{2025}\natexlab{}.
\newblock \showarticletitle{EgoLife: Towards Egocentric Life Assistant}.
\newblock \bibinfo{journal}{\emph{arXiv preprint arXiv:2503.03803}} (\bibinfo{year}{2025}).
\newblock


\bibitem[Yang et~al\mbox{.}(2024)]%
        {yang2024alignment}
\bibfield{author}{\bibinfo{person}{Yang Yang}, \bibinfo{person}{Jinyi Guo}, \bibinfo{person}{Guangyu Li}, \bibinfo{person}{Lanyu Li}, \bibinfo{person}{Wenjie Li}, {and} \bibinfo{person}{Jian Yang}.} \bibinfo{year}{2024}\natexlab{}.
\newblock \showarticletitle{Alignment efficient image-sentence retrieval considering transferable cross-modal representation learning}.
\newblock \bibinfo{journal}{\emph{Frontiers of Computer Science}} \bibinfo{volume}{18}, \bibinfo{number}{1} (\bibinfo{year}{2024}), \bibinfo{pages}{181335}.
\newblock


\bibitem[Yin et~al\mbox{.}(2024b)]%
        {yin2024tri}
\bibfield{author}{\bibinfo{person}{Kangning Yin}, \bibinfo{person}{Shihao Zou}, \bibinfo{person}{Yuxuan Ge}, {and} \bibinfo{person}{Zheng Tian}.} \bibinfo{year}{2024}\natexlab{b}.
\newblock \showarticletitle{Tri-modal motion retrieval by learning a joint embedding space}. In \bibinfo{booktitle}{\emph{Proceedings of the IEEE/CVF Conference on Computer Vision and Pattern Recognition}}. \bibinfo{pages}{1596--1605}.
\newblock


\bibitem[Yin et~al\mbox{.}(2025)]%
        {yin2025distributional}
\bibfield{author}{\bibinfo{person}{Wenzhe Yin}, \bibinfo{person}{Zehao Xiao}, \bibinfo{person}{Pan Zhou}, \bibinfo{person}{Shujian Yu}, \bibinfo{person}{Jiayi Shen}, \bibinfo{person}{Jan-Jakob Sonke}, {and} \bibinfo{person}{Efstratios Gavves}.} \bibinfo{year}{2025}\natexlab{}.
\newblock \showarticletitle{Distributional Vision-Language Alignment by Cauchy-Schwarz Divergence}.
\newblock \bibinfo{journal}{\emph{arXiv preprint arXiv:2502.17028}} (\bibinfo{year}{2025}).
\newblock


\bibitem[Yin et~al\mbox{.}(2024a)]%
        {yin2024domain}
\bibfield{author}{\bibinfo{person}{Wenzhe Yin}, \bibinfo{person}{Shujian Yu}, \bibinfo{person}{Yicong Lin}, \bibinfo{person}{Jie Liu}, \bibinfo{person}{Jan-Jakob Sonke}, {and} \bibinfo{person}{Stratis Gavves}.} \bibinfo{year}{2024}\natexlab{a}.
\newblock \showarticletitle{Domain Adaptation with Cauchy-Schwarz Divergence}. In \bibinfo{booktitle}{\emph{The 40th Conference on Uncertainty in Artificial Intelligence}}.
\newblock


\bibitem[Young et~al\mbox{.}(2014)]%
        {young2014image}
\bibfield{author}{\bibinfo{person}{Peter Young}, \bibinfo{person}{Alice Lai}, \bibinfo{person}{Micah Hodosh}, {and} \bibinfo{person}{Julia Hockenmaier}.} \bibinfo{year}{2014}\natexlab{}.
\newblock \showarticletitle{From image descriptions to visual denotations: New similarity metrics for semantic inference over event descriptions}.
\newblock \bibinfo{journal}{\emph{Transactions of the association for computational linguistics}}  \bibinfo{volume}{2} (\bibinfo{year}{2014}), \bibinfo{pages}{67--78}.
\newblock


\bibitem[Yu et~al\mbox{.}(2025)]%
        {yu2025conditional}
\bibfield{author}{\bibinfo{person}{Shujian Yu}, \bibinfo{person}{Hongming Li}, \bibinfo{person}{Sigurd L{\o}kse}, \bibinfo{person}{Robert Jenssen}, {and} \bibinfo{person}{Jos{\'e}~C Pr{\'\i}ncipe}.} \bibinfo{year}{2025}\natexlab{}.
\newblock \showarticletitle{The conditional cauchy-schwarz divergence with applications to time-series data and sequential decision making}.
\newblock \bibinfo{journal}{\emph{IEEE Transactions on Pattern Analysis and Machine Intelligence}} (\bibinfo{year}{2025}).
\newblock


\bibitem[Yu et~al\mbox{.}(2024)]%
        {yu2024cauchy}
\bibfield{author}{\bibinfo{person}{Shujian Yu}, \bibinfo{person}{Xi Yu}, \bibinfo{person}{Sigurd L{\o}kse}, \bibinfo{person}{Robert Jenssen}, {and} \bibinfo{person}{Jose~C Principe}.} \bibinfo{year}{2024}\natexlab{}.
\newblock \showarticletitle{Cauchy-Schwarz Divergence Information Bottleneck for Regression}. In \bibinfo{booktitle}{\emph{The Twelfth International Conference on Learning Representations}}.
\newblock


\bibitem[Zhai et~al\mbox{.}(2013)]%
        {zhai2013learning}
\bibfield{author}{\bibinfo{person}{Xiaohua Zhai}, \bibinfo{person}{Yuxin Peng}, {and} \bibinfo{person}{Jianguo Xiao}.} \bibinfo{year}{2013}\natexlab{}.
\newblock \showarticletitle{Learning cross-media joint representation with sparse and semisupervised regularization}.
\newblock \bibinfo{journal}{\emph{IEEE Transactions on Circuits and Systems for Video Technology}} \bibinfo{volume}{24}, \bibinfo{number}{6} (\bibinfo{year}{2013}), \bibinfo{pages}{965--978}.
\newblock


\bibitem[Zhan et~al\mbox{.}(2020)]%
        {zhan2020supervised}
\bibfield{author}{\bibinfo{person}{Yu-Wei Zhan}, \bibinfo{person}{Xin Luo}, \bibinfo{person}{Yongxin Wang}, {and} \bibinfo{person}{Xin-Shun Xu}.} \bibinfo{year}{2020}\natexlab{}.
\newblock \showarticletitle{Supervised hierarchical deep hashing for cross-modal retrieval}. In \bibinfo{booktitle}{\emph{Proceedings of the 28th ACM International Conference on Multimedia}}. \bibinfo{pages}{3386--3394}.
\newblock


\bibitem[Zhang et~al\mbox{.}(2018)]%
        {zhang2018unsupervised}
\bibfield{author}{\bibinfo{person}{Jian Zhang}, \bibinfo{person}{Yuxin Peng}, {and} \bibinfo{person}{Mingkuan Yuan}.} \bibinfo{year}{2018}\natexlab{}.
\newblock \showarticletitle{Unsupervised generative adversarial cross-modal hashing}. In \bibinfo{booktitle}{\emph{Proceedings of the AAAI conference on artificial intelligence}}, Vol.~\bibinfo{volume}{32}.
\newblock


\bibitem[Zhang et~al\mbox{.}(2025b)]%
        {zhang2025enhancing}
\bibfield{author}{\bibinfo{person}{Jiwei Zhang}, \bibinfo{person}{Yi Yu}, \bibinfo{person}{Suhua Tang}, \bibinfo{person}{GuoJun Qi}, \bibinfo{person}{Haiyuan Wu}, {and} \bibinfo{person}{Hirotaka Hachiya}.} \bibinfo{year}{2025}\natexlab{b}.
\newblock \showarticletitle{Enhancing semantic audio-visual representation learning with supervised multi-scale attention}.
\newblock \bibinfo{journal}{\emph{Pattern Analysis and Applications}} \bibinfo{volume}{28}, \bibinfo{number}{2} (\bibinfo{year}{2025}), \bibinfo{pages}{40}.
\newblock


\bibitem[Zhang et~al\mbox{.}(2025a)]%
        {zhang2025composed}
\bibfield{author}{\bibinfo{person}{Kun Zhang}, \bibinfo{person}{Jingyu Li}, \bibinfo{person}{Zhe Li}, {and} \bibinfo{person}{Jingjing Zhang}.} \bibinfo{year}{2025}\natexlab{a}.
\newblock \showarticletitle{Composed Multi-modal Retrieval: A Survey of Approaches and Applications}.
\newblock \bibinfo{journal}{\emph{arXiv preprint arXiv:2503.01334}} (\bibinfo{year}{2025}).
\newblock


\bibitem[Zhang and Lu(2018)]%
        {zhang2018deep}
\bibfield{author}{\bibinfo{person}{Ying Zhang} {and} \bibinfo{person}{Huchuan Lu}.} \bibinfo{year}{2018}\natexlab{}.
\newblock \showarticletitle{Deep cross-modal projection learning for image-text matching}. In \bibinfo{booktitle}{\emph{Proceedings of the European conference on computer vision (ECCV)}}. \bibinfo{pages}{686--701}.
\newblock


\bibitem[Zhen et~al\mbox{.}(2019)]%
        {zhen2019deep}
\bibfield{author}{\bibinfo{person}{Liangli Zhen}, \bibinfo{person}{Peng Hu}, \bibinfo{person}{Xu Wang}, {and} \bibinfo{person}{Dezhong Peng}.} \bibinfo{year}{2019}\natexlab{}.
\newblock \showarticletitle{Deep supervised cross-modal retrieval}. In \bibinfo{booktitle}{\emph{Proceedings of the IEEE/CVF conference on computer vision and pattern recognition}}. \bibinfo{pages}{10394--10403}.
\newblock


\bibitem[Zheng et~al\mbox{.}(2025)]%
        {zheng2025anatomy}
\bibfield{author}{\bibinfo{person}{Meng Zheng}, \bibinfo{person}{Jiajin Zhang}, \bibinfo{person}{Benjamin Planche}, \bibinfo{person}{Zhongpai Gao}, \bibinfo{person}{Terrence Chen}, {and} \bibinfo{person}{Ziyan Wu}.} \bibinfo{year}{2025}\natexlab{}.
\newblock \showarticletitle{Anatomy-Aware Conditional Image-Text Retrieval}.
\newblock \bibinfo{journal}{\emph{arXiv preprint arXiv:2503.07456}} (\bibinfo{year}{2025}).
\newblock


\bibitem[Zhi et~al\mbox{.}(2020)]%
        {zhi2020cross}
\bibfield{author}{\bibinfo{person}{Tao Zhi}, \bibinfo{person}{Yingchun Fan}, {and} \bibinfo{person}{Hong Han}.} \bibinfo{year}{2020}\natexlab{}.
\newblock \showarticletitle{Cross-Modal Retrieval via Similarity-Preserving Learning and Semantic Average Embedding}.
\newblock \bibinfo{journal}{\emph{IEEE Access}}  \bibinfo{volume}{8} (\bibinfo{year}{2020}), \bibinfo{pages}{223918--223930}.
\newblock


\end{thebibliography}




\clearpage
\onecolumn
\appendix


\section{Algorithms and Implementation Details}

This section provides details regarding the experimental setup used for the comparisons and the methods we employed, including the hyperparameters for the baseline methods (JFSE~\cite{xu2020joint} and LAVIMO~\cite{yin2024tri}) as well as the structure of the networks used in these methods.

\subsection{Bi-modality JFSE method with CS divergence}
We proposed Algorithm~\ref{alg:cs-bimodal}, which implements the bi-modal alignment framework described in Section~3.2, where Cauchy-Schwarz (CS) divergence~\cite{principe2000information, yu2025conditional,yu2024cauchy} replaces the traditional KL divergence~\cite{kullback1951information} in the Cross-Modal Projection Matching (CMPM) loss~\cite{zhang2018deep}. This algorithm leverages CS divergence's closed-form computation and symmetry, enabling numerically stable pairwise modality alignment. It projects features from one modality to another using temperature-scaled cosine similarities followed by softmax normalization. The symmetric structure of CS divergence is explicitly preserved via bidirectional loss computation ($\mathcal{L}^{\text{true}}_{\text{i2t}} + \mathcal{L}^{\text{true}}_{\text{t2i}}$), providing a balanced alignment between image and text embeddings. Implementation-wise, the scalar projection step mirrors the standard cosine similarity calculation, and the loss term directly follows the definition of CS divergence applied to probability distributions obtained from normalized projections.

\begin{algorithm}
\caption{Bi-Modal CMPM Alignment using Cauchy-Schwarz Divergence}
\label{alg:cs-bimodal}
\begin{algorithmic}[1]
\State \textbf{Input:} Normalized image features $\{ \mathbf{v}_i \}_{i=1}^{n}$, normalized text features $\{ \mathbf{t}_j \}_{j=1}^{n}$, binary label matrix $\mathbf{Y} = [y_{ij}] \in \{0,1\}^{n \times n}$, temperature $\tau$
\State \textbf{Output:} Average Cauchy-Schwarz divergence loss $\mathcal{L}_{\mathrm{CS}} / n$

\State Initialize $\mathcal{L}_{\mathrm{CS}} \gets 0$
\State Set temperature parameter $\tau > 0$

\For{$i = 1$ to $n$}
    \State \Comment{Step 1: Compute softmax-based projection distribution $p_{ij}$}
    \For{$j = 1$ to $n$}
        \State $s_{ij} \gets \cos(\mathbf{v}_i, \mathbf{t}_j) / \tau$
    \EndFor
    \State $p_{ij} \gets \frac{\exp(s_{ij})}{\sum_{k=1}^{n} \exp(s_{ik})}$ \Comment{Softmax over similarities}

    \State \Comment{Step 2: Normalize ground-truth matching distribution $q_{ij}$}
    \For{$j = 1$ to $n$}
        \State $q_{ij} \gets \frac{y_{ij}}{\sum_{k=1}^{n} y_{ik}}$
    \EndFor

    \State \Comment{Step 3: Compute CS divergence for sample $i$}
    \State $numerator \gets \sum_{j=1}^{n} p_{ij} \cdot q_{ij}$
    \State $denominator \gets \sqrt{\sum_{j=1}^{n} p_{ij}^2} \cdot \sqrt{\sum_{j=1}^{n} q_{ij}^2}$
    \State $\mathcal{L}_{\mathrm{CS}} \gets \mathcal{L}_{\mathrm{CS}} - \log \left( \frac{numerator}{denominator} \right)$
\EndFor

\State \Return $\mathcal{L}_{\mathrm{CS}} / n$
\end{algorithmic}
\end{algorithm}

As the core subroutine for bi-modal alignment, Algorithm~\ref{alg:jfse-cs} implements the Cauchy-Schwarz divergence-based loss function described in Section~3.2. The implementation follows the theoretical formulation of CS divergence by computing the numerator as the expectation of joint probabilities derived from projected distributions. At the same time, the denominator captures the product of their normalized magnitudes. A logarithmic transformation is applied to preserve the scale-invariant property, making the loss robust to value shifts in the input similarity space. This modular and symmetric loss computation enables plug-and-play integration into broader bi-modal retrieval frameworks and supports stable training even under synthesized feature conditions. The algorithm integrates this loss into the JFSE architecture, maintaining the original adversarial training pipeline for sample generation and applying the CS-based alignment loss to real and generated feature pairs. This validates our claim in Section~3.4 regarding the general applicability of CS divergence to supervised feature alignment tasks.

\begin{algorithm}
\caption{Joint Feature Synthesis and Embedding (JFSE) with Cauchy-Schwarz Divergence}
\label{alg:jfse-cs}
\begin{algorithmic}[1]
\State \textbf{Input:} Source set $O_S = \{(v_i, t_i, y_i)\}_{i=1}^{n}$, class embeddings $\{c_i\}_{i=1}^{n}$, learning rate $\eta$, balance weights $\lambda_{\mathrm{adv}}, \lambda_{\mathrm{cyc}}$
\State \textbf{Output:} Trained regressors $R_v$, $R_t$

\State Initialize generators $G_v, G_t$, discriminators $D_v, D_t$
\State Initialize regressors $R_v, R_t$, modality discriminator $D_m$
\For{iteration = 1 to \textit{max\_iterations}}
    \State Sample mini-batch $\{(v_i, t_i, y_i, c_i)\}_{i=1}^B$ from $O_S$
    \For{each modality $m \in \{\mathrm{image}, \mathrm{text}\}$}
        \If{$m$ is image}
            \State Sample noise $z_v \sim \mathcal{N}(0, I)$
            \State Generate fake image feature $\tilde{v}_i = G_v(z_v, c_i)$
            \State Compute $L^{\mathrm{cWGAN}}_v = \mathrm{cWGAN\_loss}(D_v, v_i, \tilde{v}_i, c_i)$
        \Else
            \State Sample noise $z_t \sim \mathcal{N}(0, I)$
            \State Generate fake text feature $\tilde{t}_i = G_t(z_t, c_i)$
            \State Compute $L^{\mathrm{cWGAN}}_t = \mathrm{cWGAN\_loss}(D_t, t_i, \tilde{t}_i, c_i)$
        \EndIf
    \EndFor

    \State Obtain embeddings from regressors:
    
    $
    e^v_i = R_v(v_i), \quad e^t_i = R_t(t_i), \quad \tilde{e}^v_i = R_v(\tilde{v}_i), \quad \tilde{e}^t_i = R_t(\tilde{t}_i)
    $

    \State Compute bi-modal alignment loss using \textbf{Algorithm~\ref{alg:cs-bimodal}}:
    \State $L^{\text{true}}_{\mathrm{CMPM-CS}} = \mathrm{BiModalCS}(e^v_i, e^t_i, y_i)$
    \State $L^{\text{syn}}_{\mathrm{CMPM-CS}} = \mathrm{BiModalCS}(\tilde{e}^v_i, \tilde{e}^t_i, y_i)$
    \State $L_{\mathrm{CMPM-CS}} = L^{\text{true}}_{\mathrm{CMPM-CS}} + L^{\text{syn}}_{\mathrm{CMPM-CS}}$

    \State Compute adversarial loss base $(D_m, e^v_i, e^t_i, \tilde{e}^v_i, \tilde{e}^t_i)$ follow JFSE's Eq.~18

    \State Compute cycle consistency loss as:
    $
    L_{\mathrm{CYC}} = \mathbb{E}\left[\|c_i - R_v(G_v(c_i, z_v))\|_2^2 + \|c_i - R_t(G_t(c_i, z_t))\|_2^2\right] + \mathbb{E}\left[\|c_i - R_v(v_i)\|_2^2 + \|c_i - R_t(t_i)\|_2^2\right]
    $

    \State Total loss for regressors:
    $
    L_R = L_{\mathrm{CMPM-CS}} + \lambda_{\mathrm{adv}} L_{\mathrm{ADV}} + \lambda_{\mathrm{cyc}} L_{\mathrm{CYC}}
    $

    \State Update parameters with gradient descent:
    $
    G_v, G_t \leftarrow G_v, G_t - \eta \nabla_{G_v, G_t} (L^{\mathrm{cWGAN}}_v + L^{\mathrm{cWGAN}}_t)
    $
    $
    D_v, D_t \leftarrow D_v, D_t - \eta \nabla_{D_v, D_t} (L^{\mathrm{cWGAN}}_v + L^{\mathrm{cWGAN}}_t)
    $
    $
    R_v, R_t \leftarrow R_v, R_t - \eta \nabla_{R_v, R_t} (L_R), \quad D_m \leftarrow D_m - \eta \nabla_{D_m} (L_{\mathrm{ADV}})
    $
\EndFor
\end{algorithmic}
\end{algorithm}

The JFSE framework employs a dual-branch architecture for processing image and text modalities.  It utilizes a VGG-19 network~\cite{simonyan2014very} pre-trained on ImageNet for visual representation, extracting 4096-dimensional feature vectors.  Textual inputs are processed through a pre-trained Word2Vec~\cite{mikolov2013efficient} model, generating 300-dimensional embeddings.  These heterogeneous features are then projected into a shared semantic space through a two-layer fully connected network.  The first transformation layer reduces the image features from 4096 to 1024 dimensions, while the second layer further compresses both modalities to a 512-dimensional unified representation, with ReLU activation functions applied at both stages to introduce non-linearity.

The JFSE method, designed for by-modal alignment, employs a learning rate of $0.0001$ and utilizes Adam optimizer with momentum parameters $\beta_1 = 0.9$ and $\beta_2 = 0.999$ to facilitate stable gradient updates. Training is carried out with a batch size of $128$, complemented by a weight decay of $10^{-5}$ to mitigate overfitting. The model undergoes training for a maximum of $100$ epochs, with an early stopping mechanism activated if the mean average precision (MAP) fails to improve over a period of $20$ consecutive epochs. To ensure effective modality alignment, the Cauchy-Schwarz (CS) divergence loss is applied during the projection matching step, quantifying the discrepancy between the two modalities. 

\subsection{Tri-modal (LAVIMO) method with GCS Divergence}
We proposed Algorithm \ref{alg:lavimo-cs}, which implements the Generalized CS divergence framework from Section~3.3. The algorithm's structure reflects our methodological innovation in joint modality alignment - where traditional pairwise methods would require three separate divergence computations, our GCS formulation achieves simultaneous alignment through a single tensor operation. The denominator's geometric mean of $M$-th order norms realizes the normalization strategy proposed in Definition~3.2, while the numerator's product term captures the global agreement metric central to our multi-modal alignment theory. The adversarial and reconstruction loss components maintain consistency with the baseline LAVIMO architecture while benefiting from GCS's enhanced alignment properties discussed in Section~4.3's experiments.

Additionally, the Algorithm \ref{a4} provides the computational backbone for tri-modal alignment, implementing the Generalized CS divergence mathematics developed in Section~3.3.1. The procedure exactly follows our extension of the CS inequality to multiple distributions through the generalized Hölder's inequality framework~\cite{qiang2011generalizations,finner1992generalization} in Eq.~8. The implementation highlights two key methodological contributions: 1) the joint probability product in the numerator evaluates global agreement across all modalities simultaneously. 2) the geometric mean normalization in the denominator maintains the divergence's symmetric properties while accommodating multiple distributions. The iterative computation across modality permutations (1→2→3→1 and reverse) ensures comprehensive alignment as theoretically justified in Eq.~15.

In comparison, the LAVIMO system extends this paradigm to accommodate four distinct modalities: visual, textual, auditory, and motion data.  Visual features are extracted using CLIP's~\cite{ramesh2021zero} vision encoder, producing 512-dimensional embeddings, while textual processing leverages DistilBERT~\cite{sanh2019distilbert} to generate 768-dimensional representations. The auditory modality employs Wav2Vec2.0~\cite{baevski2020wav2vec}, and motion data is processed through MotionClip~\cite{tevet2022motionclip}, both yielding 512-dimensional feature vectors. The fusion architecture consists of a three-layer projection network, where the initial layer concatenates and transforms all modalities from their combined 2304-dimensional input (512 + 768 + 512 + 512) to a 1024-dimensional intermediate representation. A subsequent layer reduces this to the final 512-dimensional shared embedding space, with ReLU activations ensuring non-linear transformation capabilities at each stage. 

\begin{algorithm}[H]
\caption{Tri-Modal Alignment using Generalized Cauchy-Schwarz Divergence (LAVIMO)}
\label{alg:lavimo-cs}
\begin{algorithmic}[1]
\State \textbf{Input:} Normalized image features $\{ \mathbf{v}_i \}_{i=1}^{n}$, text features $\{ \mathbf{t}_i \}_{i=1}^{n}$, audio features $\{ \mathbf{a}_i \}_{i=1}^{n}$, label matrix $\mathbf{Y} \in \{0,1\}^{n \times n}$, temperature $\tau$
\State \textbf{Output:} Average Generalized Cauchy-Schwarz divergence loss $\mathcal{L}_{\mathrm{GCS}} / n$

\State Initialize total loss $\mathcal{L}_{\mathrm{GCS}} \gets 0$

\State Compute projection distributions:
$
P_{AB}[i,j] = \frac{\exp(\cos(\mathbf{v}_i, \mathbf{t}_j)/\tau)}{\sum_{k=1}^{n} \exp(\cos(\mathbf{v}_i, \mathbf{t}_k)/\tau)} \quad \text{// image → text}
$

$
P_{BC}[i,j] = \frac{\exp(\cos(\mathbf{t}_i, \mathbf{a}_j)/\tau)}{\sum_{k=1}^{n} \exp(\cos(\mathbf{t}_i, \mathbf{a}_k)/\tau)} \quad \text{// text → audio}
$
$
P_{CA}[i,j] = \frac{\exp(\cos(\mathbf{a}_i, \mathbf{v}_j)/\tau)}{\sum_{k=1}^{n} \exp(\cos(\mathbf{a}_i, \mathbf{v}_k)/\tau)} \quad \text{// audio → image}
$

\For{$i = 1$ to $n$}
    \State $p_{AB}^{(i)} \gets P_{AB}[i]$, $p_{BC}^{(i)} \gets P_{BC}[i]$, $p_{CA}^{(i)} \gets P_{CA}[i]$
    
    \State \Comment{Compute numerator and denominator of GCS divergence}
    \State $numerator \gets \sum_{j=1}^{n} p_{AB}^{(i)}[j] \cdot p_{BC}^{(i)}[j] \cdot p_{CA}^{(i)}[j]$
    \State $norm_{AB} \gets \left( \sum_{j=1}^{n} (p_{AB}^{(i)}[j])^4 \right)^{1/4}$
    \State $norm_{BC} \gets \left( \sum_{j=1}^{n} (p_{BC}^{(i)}[j])^4 \right)^{1/4}$
    \State $norm_{CA} \gets \left( \sum_{j=1}^{n} (p_{CA}^{(i)}[j])^4 \right)^{1/4}$
    \State $denominator \gets norm_{AB} \cdot norm_{BC} \cdot norm_{CA}$
    
    \State $gcs_i \gets -\log\left( \frac{numerator}{denominator + \epsilon} \right)$ \Comment{$\epsilon$ avoids division by zero}
    \State $\mathcal{L}_{\mathrm{GCS}} \gets \mathcal{L}_{\mathrm{GCS}} + gcs_i$
\EndFor

\State \Return $\mathcal{L}_{\mathrm{GCS}} / n$
\end{algorithmic}
\end{algorithm}

\begin{algorithm}
\caption{LAVIMO with GCS Divergence for Tri-Modal Learning (Video, Text, and Motion)}
\label{a4}
\begin{algorithmic}[1]
\State \textbf{Input:} Video features \( \{v_i\} \), Text features \( \{t_i\} \), Motion features \( \{m_i\} \), Ground truth matrix \( Q \), learning rate \( \eta \), balance weight \( \lambda \)
\State \textbf{Output:} Trained encoders \( E_v, E_t, E_m \) for video, text, and motion

\State Initialize encoders \( E_v, E_t, E_m \) for each modality
\For{iteration = 1 \textbf{to} max\_iterations}
    \State Sample mini-batch \( \{(v_i, t_i, m_i)\}_{i=1}^B \)
    \State Compute embeddings:
    
    $
    e_v = E_v(v_i), \quad e_t = E_t(t_i), \quad e_m = E_m(m_i)
    $
    \State Compute projection probabilities via cosine+softmax:
    $
    P_{vt} = \texttt{softmax}(\cos(e_v, e_t)), \quad P_{tm} = \texttt{softmax}(\cos(e_t, e_m)), \quad P_{mv} = \texttt{softmax}(\cos(e_m, e_v))
    $
    \State Compute GCS-based alignment loss using \textbf{Algorithm~\ref{alg:lavimo-cs}}:
    $
    L_{\text{align}} = \mathrm{GCS\_Divergence}(P_{vt}, P_{tm}, P_{mv})
    $
    \State Compute attention-based modality interaction loss:
    \State \quad $A_{vt} = \mathrm{Attention}(e_v \rightarrow e_t)$
    \State \quad $A_{tm} = \mathrm{Attention}(e_t \rightarrow e_m)$
    \State \quad $A_{mv} = \mathrm{Attention}(e_m \rightarrow e_v)$
    \State \quad $L_{\text{attn}} = \sum_{\text{pairs}} \|A - \mathrm{detach}(\text{SoftSim})\|^2$
    \Comment{Attention should match similarity supervision}
    \State Compute reconstruction loss:
    \State \quad $L_{\text{recon}} = \sum \|e_v - \hat{e}_v\|^2 + \|e_t - \hat{e}_t\|^2 + \|e_m - \hat{e}_m\|^2$ \\
    \hfill (e.g., via shared decoder or cyclic regression)
    \State Total loss:
    $
    L_{\text{total}} = L_{\text{align}} + \lambda_{\text{attn}} L_{\text{attn}} + \lambda_{\text{recon}} L_{\text{recon}}
    $
    \State Update all encoders:
    $
    E_v, E_t, E_m \leftarrow E_v, E_t, E_m - \eta \cdot \nabla_{E_v, E_t, E_m}(L_{\text{total}})
    $
\EndFor
\end{algorithmic}
\end{algorithm}

The LAVIMO three modalities while maintaining identical foundational hyperparameters, including learning rate, optimizer configuration, batch size, weight decay, and early stopping criteria. The divergence from the JFSE method arises in the adoption of the generalized Cauchy-Schwarz (GCS) divergence loss, which measures joint alignment across all three modalities through a tensor-based interaction framework. Additionally, adversarial loss is incorporated to further refine feature alignment, while reconstruction loss, inclusive of cycle consistency constraints, ensures modality-specific feature preservation throughout the training process.

\section{Theoretical Properties of Generalized Cauchy-Schwarz Divergence}
In this section, we present rigorous proofs for the fundamental properties of the Generalized Cauchy-Schwarz  (GCS) divergence defined in Eq.~13. Specifically, we prove that GCS divergence satisfies the following desirable properties: \textbf{non-negativity}, \textbf{symmetry}, and \textbf{scale invariance} (also referred to as \textbf{projective invariance}). These properties collectively guarantee the validity and stability of the proposed divergence measure when applied to multiple discrete distributions.

Let $P_1, \ldots, P_M$ be a finite set of discrete distributions over common support of $K$ elements, where each $P_m$ is represented by a probability vector $p_m = \{p_{m,k}\}_{k=1}^K$ satisfying $\sum_{k=1}^K p_{m,k} = 1$.

\subsection{Non-negativity}
We first show that $D_{GCS}(P_1, \ldots, P_M) \geq 0$ always holds, with equality if and only if all distributions are proportional. From the generalized Hölder's inequality, we have:
\[
\sum_{k=1}^K \prod_{m=1}^M p_{m,k} \leq \prod_{m=1}^M \left( \sum_{k=1}^K p_{m,k}^M \right)^{\frac{1}{M}}.
\]

Taking logarithms of both sides and negating, we obtain:
\[
D_{GCS}(P_1, \ldots, P_M) = -\log \left( \frac{\sum_{k=1}^K \prod_{m=1}^M p_{m,k}}{\left( \prod_{m=1}^M \left( \sum_{k=1}^K p_{m,k}^M \right) \right)^{1/M}} \right) \geq 0.
\]

The equality holds if and only if the distributions are collinear, i.e., there exist positive constants $\{\beta_m\}$ such that $\beta_1 P_1 = \cdots = \beta_M P_M$. This ensures that the GCS divergence vanishes only when all distributions are identical up to scaling, as required by a valid divergence.

\subsection{Symmetry}
The symmetry of GCS divergence follows directly from the commutativity of the product and sum operations in the definition. Specifically, for any permutation $\pi$ of $\{1, \ldots, M\}$, we have:
\[
\mathcal{D}_{GCS}(P_1, \ldots, P_M) = \mathcal{D}_{GCS}(P_{\pi(1)}, \ldots, P_{\pi(M)}).
\]
This confirms that GCS divergence treats all input distributions equally, an essential requirement for divergence measures defined over unordered sets of distributions.

\subsection{Scale Invariance (Projective Invariance)}

We now show that GCS divergence is invariant under positive scaling of input distributions, i.e., for any set of positive scalars $\{\beta_m > 0\}$, the divergence satisfies:
\[
\mathcal{D}_{GCS}(\beta_1 P_1, \ldots, \beta_M P_M) = \mathcal{D}_{GCS}(P_1, \ldots, P_M).
\]
To see this, consider:
\[
\sum_{k=1}^K \prod_{m=1}^M (\beta_m p_{m,k}) = \left( \prod_{m=1}^M \beta_m \right) \sum_{k=1}^K \prod_{m=1}^M p_{m,k},
\]
and
\[
\prod_{m=1}^M \left( \sum_{k=1}^K (\beta_m p_{m,k})^M \right)^{1/M} = \left( \prod_{m=1}^M \beta_m \right) \prod_{m=1}^M \left( \sum_{k=1}^K p_{m,k}^M \right)^{1/M}.
\]
The multiplicative scaling terms cancel out in the numerator and denominator, leaving the divergence value unchanged. Therefore, GCS divergence is invariant to the scaling of input distributions, making it robust to normalization and suitable for applications where the scale is arbitrary or uninformative.

These properties collectively demonstrate that the GCS divergence is a \textbf{valid}, \textbf{stable}, and \textbf{geometrically meaningful} divergence measure over multiple distributions. This solid theoretical foundation supports its application in multi-distribution learning problems such as deep clustering and multi-source domain adaptation.

\subsection{Integration of CS/GCS Divergence into RONO, COXI}

To validate the flexibility of our CS-based alignment, we integrated CS/GCS into two representative retrieval methods \textbf{RONO} and \textbf{COXI} by replacing their original embedding alignment losses while preserving other architecture and training settings, as summarized in \textbf{Table~\ref{tab:RONO}}.

RONO~\cite{feng2023rono} uses a robust discriminative center loss to encourage intra-class compactness and inter-class separation, together with a multimodal gap loss. We substitute the center-based loss with CS divergence. On Flickr30K, RONO+CS raises average P@10 from 0.845 to 0.874 and RONO+GCS further improves to 0.886, with the most notable gains (+2.1\%) in audio-text retrieval (A2T/T2A).

COXI~\cite{wei2024cross} employs a cross-modal shared-proxy loss, aligning each modality via proxy points in embedding space. We replace this proxy alignment with our GCS divergence, enabling direct tri-modal alignment. COXI+GCS achieves an average P@10 of 0.823 on Flickr30K, improving over COXI+CS (0.838) and the original (0.793), with strong gains in A2T/T2A (+0.82). This demonstrates improved embedding coherence across three modalities.

All variants use identical feature extractors ResNet50 for images, DistilBERT for text, and Wav2Vec2.0 for audio with batch size = 64, learning rate = $10^{-4}$, and Adam optimizer. Results are averaged across 10 runs.
These experiments confirm that CS/GCS divergence can be seamlessly inserted into diverse retrieval architectures, yielding consistent performance gains (+2–5\% P@10) without structural changes or extra hyperparameters validating its value as a general-purpose alignment module.

\begin{table}[htbp]\setlength{\tabcolsep}{1.2pt}
	\small 
	\caption{Tri-modal retrieval performance on Flikcer30K Datasets. The MAP scores (P@K, K=10) of two original methods with their CS and GCS variants.}
	\label{tab:RONO}
	\begin{tabular}{c|ccc|ccc|ccc}
		\toprule
		\multirow{2}{*}{Method} & \multicolumn{9}{c}{Flikcer30K} \\
		\cmidrule(l{6.5em}r{0.3em}){1-4}  \cmidrule(l{2.9em}r{2.9em}){4-8}  \cmidrule(l{2.9em}r{0em}){7-10}  
		& I2T & T2I & Avg.& I2A & A2I & Avg. & A2T & T2A & Avg.  \\
		\midrule
        RONO\cite{feng2023rono}&0.695&0.702&0.699&0.243&0.181&0.212&0.811&0.878&0.845\\
        \rowcolor{gray!25} RONO+\bfseries CS&0.697&\bfseries0.710&0.704&\bfseries0.275&0.213&0.244&0.847&0.900&0.874\\ 
        \rowcolor{gray!25} RONO+\bfseries GCS&\bfseries0.709&0.707\bfseries&0.708&0.261&\bfseries0.231&\bfseries0.246&\bfseries0.849&\bfseries0.923&\bfseries0.886\\
        \midrule
        COXI\cite{wei2024cross}&0.699&0.672&0.686&0.202&0.168&0.185&0.747&0.839&0.793\\
        \rowcolor{gray!25} COXI+\bfseries CS&\bfseries0.722&0.693&\bfseries0.708&\bfseries0.251&\bfseries0.201&\bfseries0.226&0.792&0.883&0.838\\
        \rowcolor{gray!25} COXI+\bfseries GCS&0.708&\bfseries0.698&0.703&0.248&0.198&0.223&\bfseries0.811&\bfseries0.898&\bfseries0.855\\
        \midrule
		LAIMAU\cite{yin2024tri}&0.692&0.667&0.680&0.288&0.267&0.278&0.964&0.990&0.977\\
        \rowcolor{gray!25}	
        LAIMAU+\bfseries CS&\bfseries0.795&0.705&\bfseries0.750&0.305&0.328&0.317&\bfseries0.985&0.993&\bfseries0.989\\
	  \rowcolor{gray!25}
        LAIMAU+\bfseries GCS&0.705&\bfseries0.712&0.709&\bfseries0.323&\bfseries0.358&\bfseries0.341&0.971&\bfseries0.997&0.984\\
		\bottomrule
	\end{tabular}
\end{table}

\subsection{Comparison with Sum of Pairwise CS}
Unlike the sum of pairwise CS divergences, which separately aligns each modality pair (e.g., I–T, T–A, I–A) and may introduce redundancy or conflicting gradient signals, our Generalized CS (GCS) divergence jointly aligns all three modality distributions by treating them as a unified structure. This holistic formulation, grounded in Hölder’s inequality, ensures consistent and symmetric alignment across modalities without requiring exhaustive pairwise comparisons.

To further clarify the role of matching strategies, we implement three variants of circular matching: (1) clockwise matching, which sequentially aligns modality distributions in the order I→T→A→I; (2) counterclockwise matching, which uses the reversed path A→T→I→A; and (3) bidirectional mixed matching, which incorporates both directions simultaneously. While the unidirectional variants capture partial dependency, the mixed strategy fully exploits the semantic complementarity among modalities, leading to more stable optimization and better retrieval performance. As shown in Table~\ref{tab:RONO}, the bidirectional GCS strategy consistently outperforms the pairwise CS baseline, confirming the advantage of unified circular alignment.

\section{ Additional Experimental Results}
\begin{figure}[t]
    \centering
    \includegraphics[width=.9\columnwidth, trim=1pt 1cm 1pt 1pt, clip, page=3]{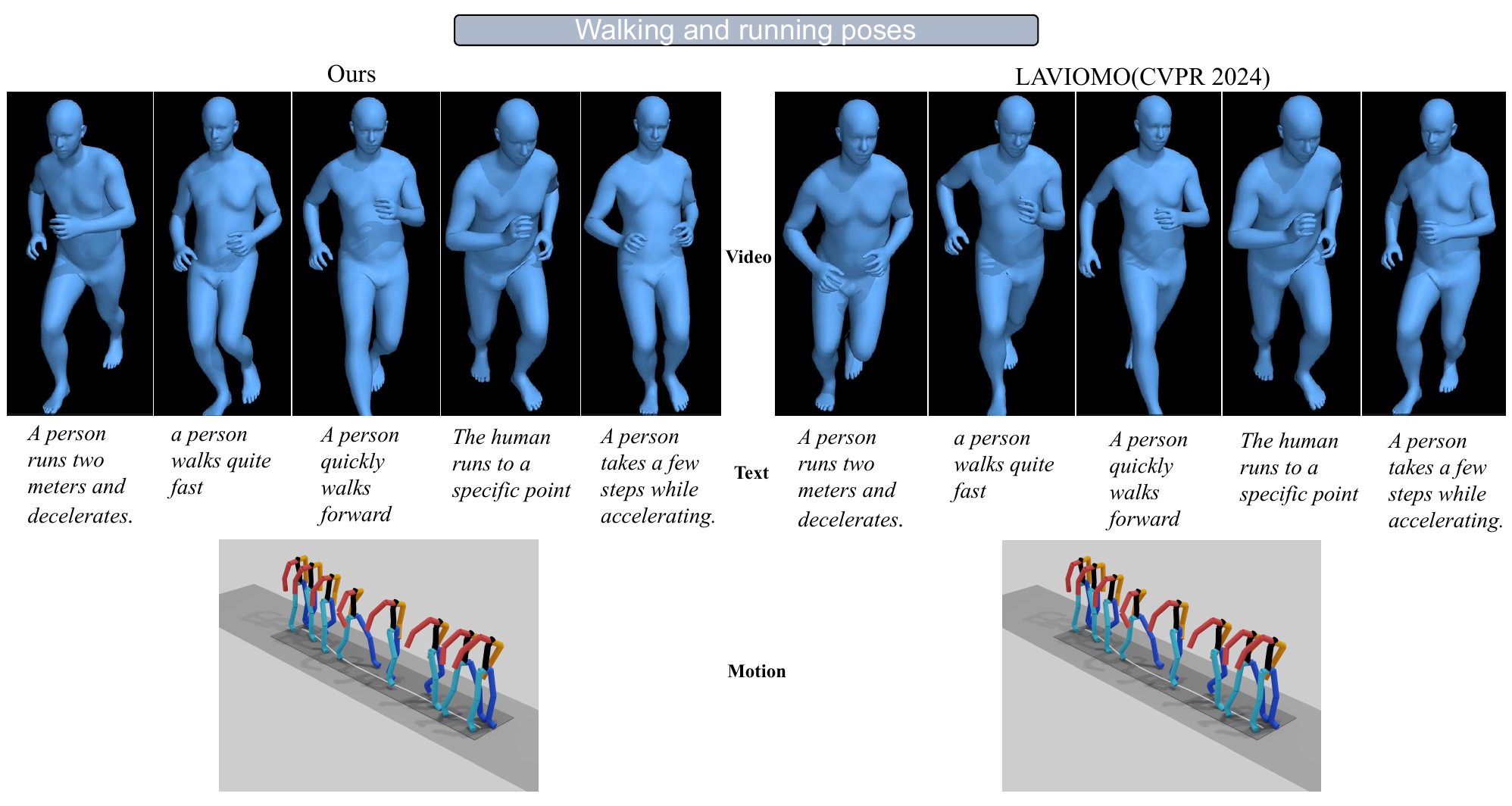}
    \caption{Comparison of video-text-motion retrieval results in the walking and running scenario.}
    \label{fig:Motion_Retrieval}
\end{figure}

To complement our quantitative evaluations, we compare the conventional Kullback-Leibler (KL) divergence and our proposed Generalized Cauchy-Schwarz (GCS) divergence in tri-modal retrieval. As shown in Figure~\ref{fig:Motion_Retrieval}, we visualize the top retrieved motion sequences for both text-to-motion and video-to-motion retrieval scenarios, using GCS (left) and KL (right, based on TMR [31]) as the alignment objectives.

\subsection{Video-Text-Motion Retrieval}

Figure~\ref{fig:Motion_Retrieval} presents a qualitative comparison of video-text-to-motion retrieval under the "walking and running" scenario. Each video query is rendered using motion-driven avatars and paired with descriptive text to form a joint query. Our GCS-based method retrieves motion sequences that more accurately reflect the underlying semantics, such as speed, posture, and intent (e.g., "quickly walks forward" or "takes a few steps while accelerating"). The retrieved motions exhibit fine-grained temporal consistency with the input.

In contrast, LAVIOMO returns more generic motions with limited variation, occasionally misaligning semantic cues such as confusing walking with running. These results highlight the effectiveness of GCS divergence in modeling subtle cross-modal interactions, particularly in scenarios that involve nuanced human movements and multi-modal inputs.

\subsection{Feature Visualization}

To gain deeper insight into the effectiveness of our divergence-based alignment methods, we provide t-SNE~\cite{van2008visualizing} visualizations of feature distributions under two-modality (image-text) and three-modality (image-text-audio) scenarios. These visualizations highlight the impact of different distribution alignment strategies on the structure of the learned shared embedding space.

\begin{figure}[t]
    \centering
    \includegraphics[width=.9\columnwidth, trim=2cm 1cm 1pt 1pt, clip, page=1]{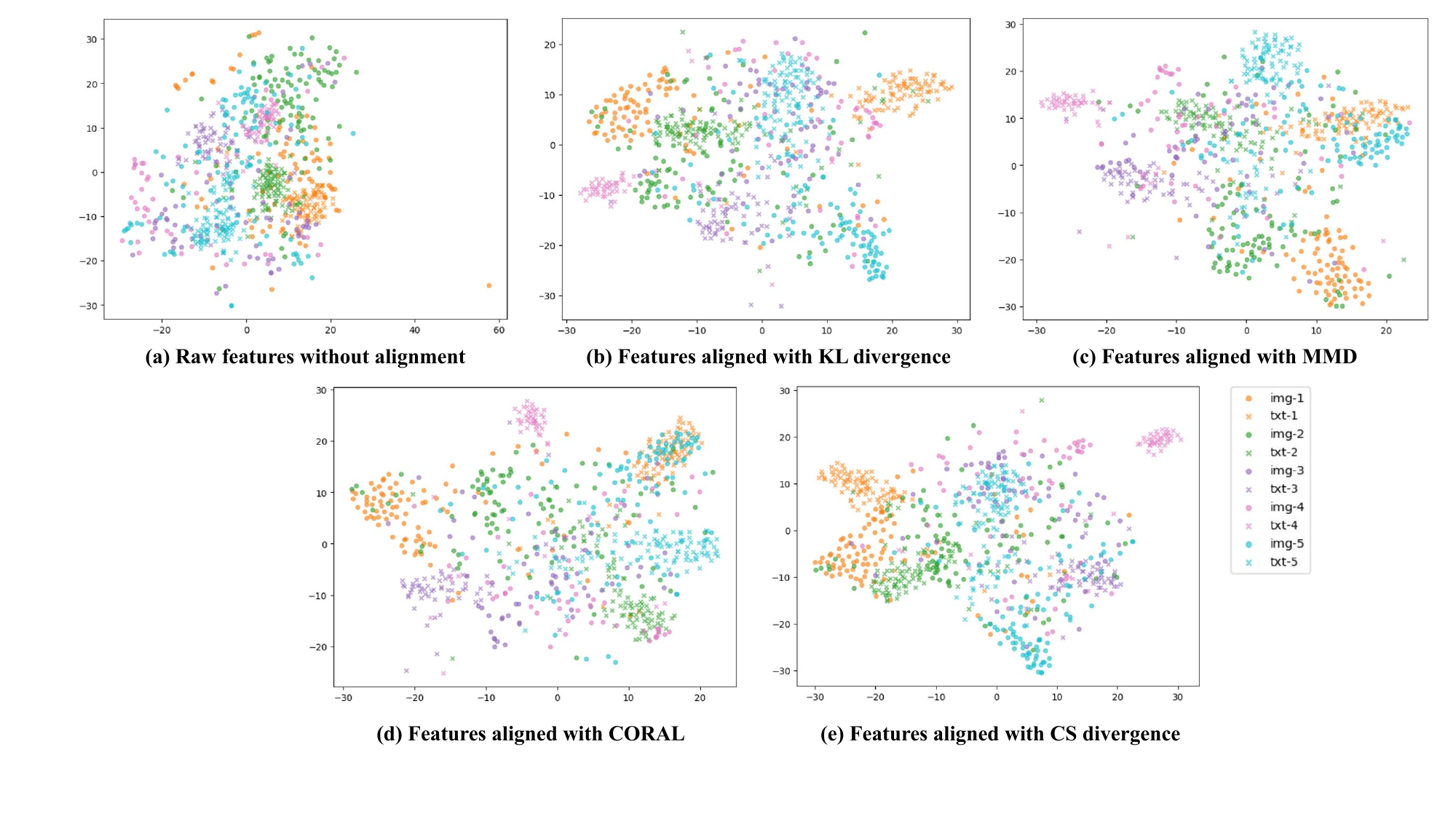}
    \caption{Visualization of image-text feature alignment under different divergence metrics.}
    \label{fig:feature_alignment}
\end{figure}
We conduct visual comparisons for two-modality alignment (Image-Text) under the JFSE framework.
Figure~\ref{fig:feature_alignment} visualizes the distribution of image and text features under different alignment strategies. In the unaligned setting (Figure~\ref{fig:feature_alignment}(a)), the two modalities form separated clusters, indicating a substantial distributional discrepancy. Applying KL divergence (Figure~\ref{fig:feature_alignment}(b)) yields partial convergence but still results in loosely coupled clusters, attributed to the asymmetry and instability of KL. The MMD-based alignment (Figure~\ref{fig:feature_alignment}(c)) improves overlap between modalities yet suffers from kernel sensitivity and limited global structure. CORAL (Figure~\ref{fig:feature_alignment}(d)) enhances modality cohesion by aligning second-order statistics but fails to capture higher-order dependencies, leading to suboptimal embeddings. In contrast, our CS-based method (Figure~\ref{fig:feature_alignment}(e)) produces compact, semantically coherent clusters, effectively reducing modality boundaries due to its symmetric and parameter-free formulation.

\begin{figure}[t]
    \centering
    \includegraphics[width=.9\columnwidth, trim=2cm 1cm 1pt 1pt, clip, page=2]{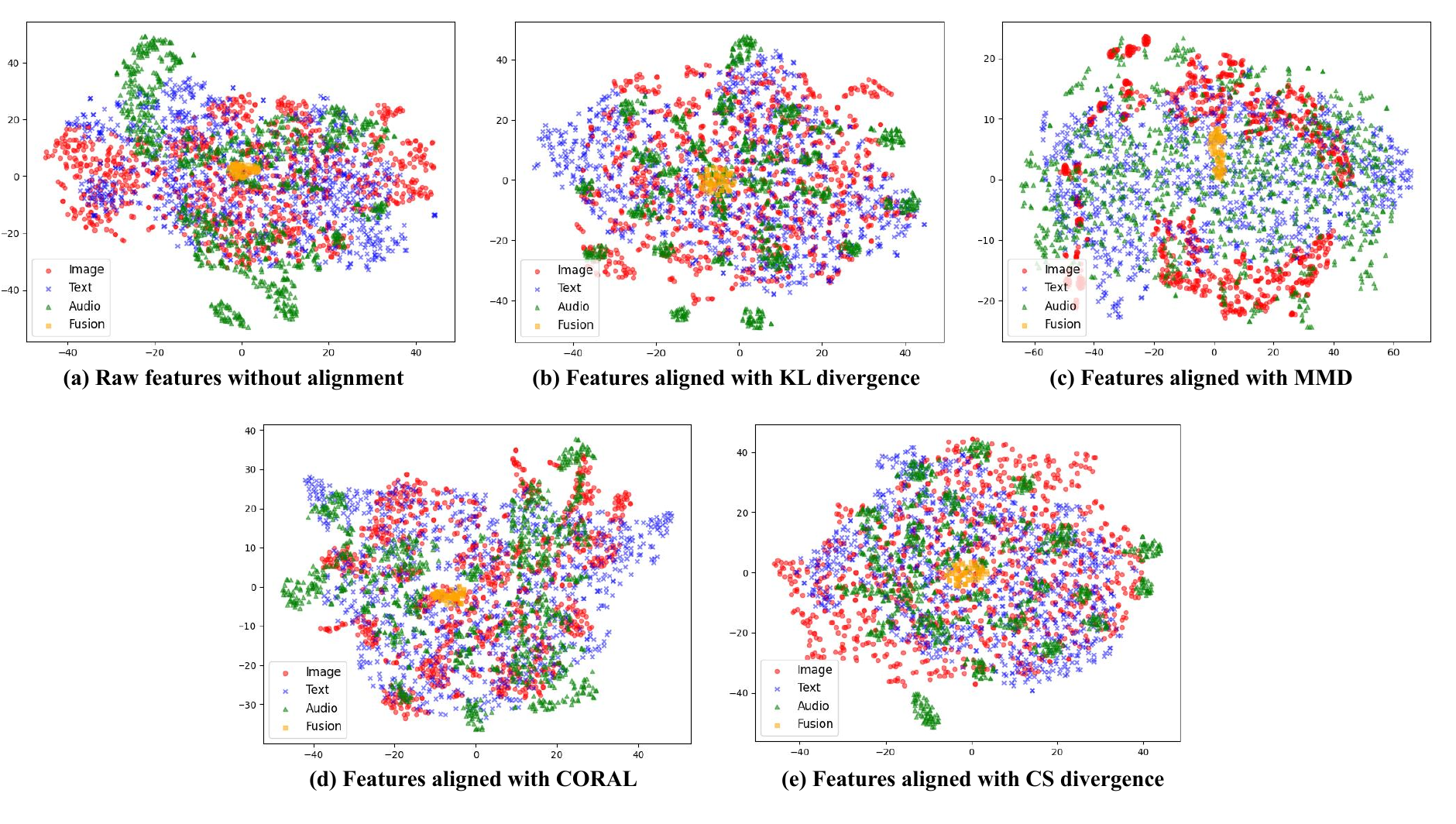}
    \caption{Visualization of image-text-audio feature alignment under different divergence metrics.}
    \label{fig:Tri_feature_alignment}
\end{figure}

We further extend this analysis to tri-modal feature alignment across image, text, and audio modalities, as shown in Figure~\ref{fig:Tri_feature_alignment}. Without alignment (Figure~\ref{fig:Tri_feature_alignment}(a)), the three modalities are distinctly separated in the latent space, lacking any shared representation. KL divergence (Figure~\ref{fig:Tri_feature_alignment}(b)) achieves limited pairwise alignment but fails to model joint modality dependencies due to its two-distribution constraint. MMD (Figure~\ref{fig:Tri_feature_alignment}(c)) introduces better feature mixing between some modalities yet struggles with kernel hyperparameter sensitivity and lacks holistic structure. CORAL (Figure~\ref{fig:Tri_feature_alignment}(d)) aligns covariance structures across modality pairs but yields fragmented clusters with weak semantic consistency. By contrast, our GCS-based approach (Figure~\ref{fig:Tri_feature_alignment}(e)) provides global alignment across all three modalities simultaneously, resulting in tight, semantically aligned clusters with clear cross-modal fusion in the shared embedding space.  

The t-SNE results reveal consistent trends across both two-modality and three-modality scenarios. CS divergence significantly improves bi-modal alignment by enforcing a symmetric and stable measure of distribution similarity. GCS divergence extends this capability to the tri-modal case, achieving holistic alignment across heterogeneous data types. In contrast, KL divergence, MMD, and CORAL either lack global alignment capacity or exhibit sensitivity to hyperparameters. These visual insights corroborate our quantitative findings, reinforcing the effectiveness and scalability of our proposed alignment framework for multi-modal retrieval.

\begin{figure}[b]
    \begin{center} 
    \includegraphics[width=1\columnwidth, trim=1.5pt 9.5cm 1pt 2cm, clip]{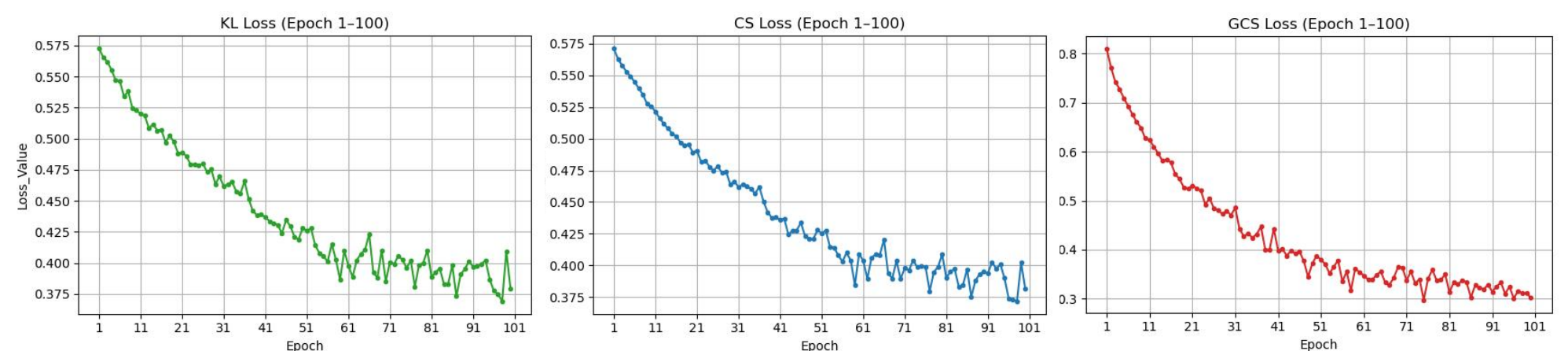}
    \caption{Training loss curves of three alignment strategies (KL, CS, and GCS) over 100 epochs. }
    \label{fig:loss}
    \end{center}
\end{figure}

\subsection{GCS Yields Smoother and More Stable Optimization}
As illustrated in Figure~\ref{fig:loss}, the KL-based alignment exhibits high initial loss and notable fluctuations in later stages, indicating unstable optimization. In contrast, the CS loss curve is smoother but still affected by local variations due to pairwise misalignment. Our proposed GCS loss shows the most stable and monotonic decline, reflecting its superior numerical stability and global coordination during training.

\end{document}